\documentclass{aa}  
\usepackage{natbib}
\usepackage{graphicx}
\usepackage{txfonts}
\usepackage{hyperref}
\usepackage{caption}
\hypersetup{colorlinks=true,allcolors=[rgb]{0,0,0.8}}
\usepackage{amsmath}
\usepackage[switch]{lineno}
\usepackage[normalem]{ulem}
\usepackage{subcaption}
\usepackage{graphicx}
\makeatletter
\renewcommand*\aa@pageof{, page \thepage{} of \pageref*{LastPage}}
\makeatother

\begin{document} 

    \title{Interior–Atmosphere Coupling on TRAPPIST-1 f, g, and h: Cryovolcanic Water Exospheres and Infrared Detectability}



\titlerunning{Water exospheres in the TRAPPIST-1 system}

  \author{E. Kleisioti
         \inst{1}\fnmsep
         \inst{2} \fnmsep
         \inst{4}
          \and
          D. Dirkx\inst{2}
          \and
          Apurva V. Oza\inst{3}
          \and
          A. Louca\inst{1}
          \and
          M. Rovira-Navarro 
          \inst{2}
          \fnmsep
          \and
          T.-M. Bründl
          \inst{2}
          \and 
          M. A. Kenworthy
          \inst{1}
          }

   \institute{Leiden Observatory, Leiden University, P.O. Box 9513, 2300 RA Leiden, The Netherlands\\
              \email{kleisioti@mail.strw.leidenuniv.nl}
         \and
             Faculty of Aerospace Engineering, TU Delft, Building 62 Kluyverweg 1, 2629 HS Delft, the Netherlands 
        \and 
             Division of Geological and Planetary Sciences, California Institute of Technology, Pasadena, CA, USA
        \and
         National Observatory of Athens, Institute of Astronomy, Astrophysics, Space Applications and Remote Sensing, 152 36 Athens, Greece}

  \abstract
   {}
   {We investigate the interior structures and cryovolcanic observability of the exoplanets TRAPPIST-1f, g, and h. Our aim is to determine which interior configurations can sustain subsurface liquid water oceans in thermal equilibrium and to assess whether the resulting cryovolcanic outgassing could be detectable with current and future observatories. } 
   {Using a layered, radially symmetric interior model that includes silicate and ice layers, we identify interior configurations in thermal equilibrium and quantify the partitioning of internal heat among the different layers and rheological parameters through a Monte Carlo analysis. We also estimate cryovolcanic water outgassing rates and assess their detectability through transmission spectroscopy, by creating synthetic transmission spectra of hydrostatic atmospheres and sputtered exospheres via radiative transfer.
}
   {We find that, for all three planets, the internal heat budget is dominated by radiogenic heating and tidal dissipation in high-pressure ice layers, with other contributions remaining minor. across the explored parameter space. 
 Thermal equilibrium solutions for the inner planets TRAPPIST-1f and g require relatively thin outer ice~I shells, implying shallow subsurface oceans, in agreement with previous work. In contrast, TRAPPIST-1h favors thicker ice shells. We show that globally distributed, non-hydrostatic sputtered exospheres require higher energy conversion efficiencies to produce detectable signals with JWST/NIRISS compared to localized plume-like outgassing configurations. Spatial localization of outgassed material in plume-like configurations substantially enhances effective line-of-sight column densities. On TRAPPIST-1f, both globally uniform and localized outgassing scenarios can produce signals above the JWST/NIRISS detectability threshold corresponding to $\sim$ 20 transits, depending on the adopted modeling assumptions.
}
 {Our interior model demonstrates that subsurface oceans are sustained over a broad range of interior configurations. Overall, our results place constraints on the observability of cryovolcanic water vapor on the TRAPPIST-1 f, g, h planets and highlight interior heat budgets and spatial distribution of outgassed material as key factors governing detectability. This framework strengthens the prospects for detecting cryovolcanically generated water vapor beyond the Solar System and motivates transmission studies of Europa-like exoplanets, that is, icy planets with subsurface oceans sustained by internal heating.}

   %


   \maketitle

   \date{Received  XX, XXXX; accepted XX, XXXX}


%

\section{Introduction}

Since the discovery of the Trappist-1 planet system \citep{gillon2017}, the system has provided an intriguing opportunity to the exoplanet community for observational and theoretical studies\citep[see e.g.][] {Krissansen-Totton_2022, Zieba_2023, Pichierri2024, schoonenberg2019}). 
Trappist-1 hosts seven known, transiting exoplanets in orbital resonance, and in almost edge-on orbits \citep{Teyssandier2022, Agol_2021}.
It is the first known system of potentially rocky, terrestial exoplanets with incident fluxes in the range of the solar system rocky planets, three of which are located within the star's habitable zone. 

The small radius of the star aids the planets' characterisation.
\citet{Agol_2021} performed an analysis of transit timing variations (TTVs) that includes all transit data from Spitzer since the discovery of the system, and provided precise mass, and radius measurements for the planets.
In addition, they observed a slight downward trend of density with orbital distance.
One possible explanation is different volatile/water content for the seven planets. 
If confirmed, the three outer planets, f,g, and h, might have between 4.5 and 6.4 wt$\%$ water assuming an Earth-like core \citep{Agol_2021}.
The latter, together with their low equilibrium temperatures, makes them good candidates for  ``cold ocean planets'' \citep{Quick_2020}.

 Cold ocean planets are ``ocean-bearing worlds with ice covered surfaces'' \citep{Quick_2020}.
 Their interior structure and properties are expected to be similar to those of the solar system icy moons \citep{Ehrenreich_2006}. 
 As such, they might harbor subsurface oceans fueled by tidal or radiogenic heating, underneath an icy shell \citep{nimmo_pappalardo2016}.
 Cold ocean planets might be in abundance in the galaxy \citep{zeng_2019}, and are significant targets for habitability and astrobiology studies. 
 Several candidate worlds of this nature have been proposed.
 For example, \citet{Quick_2023} identified 17 planet candidates that might be icy, including Trappist-1,f,g,h, according to their measured densities and equilibrium temperatures. 
Another planet, whose properties are compatible with a water-rich planet covered in an icy surface is LHS 1140 b \citep{Quick_2020, Quick_2023, Cadieux_2024}. 

 Several studies have characterized the possible water-bearing interior of Trappist-1g,f,h planets \citep[see e.g. ][]{Barr_2018, Dobos2019}.
 \citet{acuna2021} studied their putative hydrospheres by modelling their pressure-temperature profiles, and concluded that water on the planets would be in solid phases.
 The planets’ non zero orbital eccentricity and short orbital periods suggest that tidal heating might be significant in their thermal budget \citep{luger2017}. 
 \citet{Barr_2018} studied tidal heating in the Trappist-1 system, and concluded that the interior of the planets can have similar geodynamics as solar system icy moons. 
 %

Cryovolcanism is the eruption or extrusion of volatile materials, such as water, aqueous solutions, or other ices, in liquid or vapor form that would otherwise be frozen at the surface of icy planetary bodies where temperatures fall below the desorption temperature of most volatiles \citep{Geissler2015}.
If the outer Trappist-1 planets have similar geological features to the solar system icy moons, they might also exhibit cryovolcanism. 
Cryovolcanism has been detected on Enceladus \citep{porco_helfenstein2006, Dougherty2006}, Triton \citep{Croft1995}, and other icy bodies in the Solar System \citep{FAGENTS2022} and it is a source of water outgassing that contributes to the formation of exospheres among other processes such as surface sputtering. In addition cryovolcanic activity has long been hypothesized for Europa \citep{FAGENTS2000, Fagents2003}. HST Ly-$\alpha$ observations observations \citep{Roth2014} and Galileo magnetic and plasma wave signatures \citep{Jia2018} were previsouly interpreted as evidence for plume activity. A more recent analysis, however, suggests that the HST Ly-$\alpha$ observations are more consistent with an atomic hydrogen exosphere than active plumes \citep{Roth_2026}.
In the Solar System, several icy moons possess tenuous exospheres or outgassing activity, such as Europa, which has an oxygen dominated exosphere \citep{SMYTH2006510, OZA201923}, and Enceladus, which possesses 
plumes of water vapor and ice grains, containing traces of salts, silicates and organic compounds \citep{Postberg2011, porco_helfenstein2006}. These plumes feed the E-ring, a diffuse torus around Saturn which extends several satellite radii from Enceladus \citep{BAUM1981}. 

Our knowledge of the atmospheres of the Trappist-1 planets is limited, including whether or not they possess atmospheres in the first place.
The extended pre-main-sequence (PMS) phase of M-dwarf stars, like Trappist-1, makes the planets subject to atmospheric loss due to their star's prolonged period of high luminosity and XUV emission \citep{Luger_2015}.
Nevertheless, efforts have been made to observationally characterize the Trappist-1 atmospheres; 
\citet{de_Wit_2018} ruled out cloudless, hydrogen-dominated atmospheres for planets d, e and f. 
In addition, \citet{Zieba_2023} found evidence of either tenuous or no atmosphere for planet c, and the photometric secondary eclipse
observations of \citet{Greene_2023} hint that planet b is most likely a bare rock. Motivated by these results, in this work we adopt a simplified approach in which the outer TRAPPIST-1 planets are treated as essentially bare icy bodies, and we model only a tenuous, cryovolcanically supplied water exosphere, neglecting any additional background atmosphere.

Tidal models of varying complexity have been applied to study bodies and moons within the Solar System \citep{rovirra_navaro2022, Bierson_2021} and exoplanet systems \citep{Henning_2014, Shoji_2014, vanWoerkom2024}. They range from models that assume tidal parameters that are independent of the orbital period and interior structure (constant-Q approach), and more complicated multi-layered approaches, which take into account the viscoelastic behavior of each layer material properties \citep{Kleisioti2023, Rovira-Navarro_2021}.
Since the discovery of the Trappist planets, several authors have attempted to constrain their interior structure and potential dynamics assuming that the outer planets Trappist-1 f,g, and h, are ``cold ocean worlds''.
\citet{Quick_2023} obtained ice shell thicknesses, tidal heating rates and cryovolcanic column densities.
They used the constant-Q approach to assess tidal heating, which did not take the dependency of the tidal parameters on each of the planets’ specific interior structure and orbit into account.


%
\citet{Barr_2018} also modeled tidal heating in the system, however they did not use the updated mass and radius constraints of \citet{Agol_2021}, resulting in significantly higher predicted water contents.

In this work, we examine the interior structure compatible with the updated density estimates using a multilayer, viscoelastic approach and examine the potential for cryovolcanism and its remote detection. By constraining the heat generated within each layer, we can determine the range of internal structures compatible with a given ice shell thickness and assess the total internal heat budget.
We focus on the possibility that the outer planets Trappist-1 f,g, and h contain water and do not possess thick atmospheres, with the goal to assess possible thermal equilibrium states, and whether exospheres, similar to the solar system icy moons, can be detectable with JWST.
We model fully differentiated planets with an icy shell, subsurface ocean, high pressure ice layer (HPI), mantle, and a core, and obtain interior structures in thermal equilibrium.
We study the possible thermal states of an icy shell and a subsurface ocean on Trappist-1f,g,h, using thermal models that have been applied to solar system icy satellites.
To do so,  we use the planet masses and radii of \citet{Agol_2021}, together with a tidal model that captures the behavior of material properties as a function of the orbital period.
In Section \ref{sec:methods} we describe our interior, thermal and tidal models.
In Section \ref{interior_results} we present the partitioning of tidal dissipation within the planet interior, and the resulting interior constraints for Trappist-1f. 
Sections \ref{sec:cryovolcanism_methods} and \ref{sec:transmission_spectra} focus on the resulting cryovolcanic activity and its predicted transmission spectroscopy signatures.
Finally, Section \ref{sec:discussion} compares our results with previous studies and discusses whether interior structures in thermal equilibrium can produce cryovolcanic signatures detectable with JWST.
%

\section{Interior structure and tidal model}\label{sec:methods}
We assume a fully differentiated body, with radially homogeneous layer properties that consist of an iron core, a silicate mantle, a high pressure ice (HPI) layer that accounts for the possible presence of the water ice polymorphs II through VII \citep[e.g. ][]{Jaccard_1976}, a liquid, subsurface ocean and an ice I shell  with densities $\rho_c$, $\rho_m$, $\rho_{HPI}$, $\rho_{ocean}$, and $\rho_{ice,I}$ \citep{Dobos2019, Barr_2018}. For our work, we require a model to determine feasible interior structures of the planets that are compatible with the mass and radius values (see Table \ref{table:planet_parameters}), under the assumption of thermal equilibrium. Here we describe the thermal model (Section \ref{sec:heat_tranfer_model}) used to calculate the equilibrium states of the ice I shell, and the tidal model (Section \ref{tidal_response_model}) to determine the heating in the dissipative layers of the planets' interior. We describe the planets' thermal equilibrium model in Section \ref{thermal_equilibrium}.
\subsection{Heat transfer model}\label{sec:heat_tranfer_model}
For the heat transfer model of the ice I shell, we follow the approach of \citet{sohl_2003}, who modelled the thermal structure of Titan's ice shell. The ice I shell is divided into a conductive outer layer and a convective sub-layer, as is typically modeled for solar system icy moons \citep[see e.g. ][]{HUSSMANN2002143, green_2021}. The heat that is transported through the conductive stagnant lid layer is:

\begin{equation} \label{eq:conduction}
\frac{Q_{cond}}{4 \pi R_P^2} = k\frac{(T_{top} - T_S)}{D_{cond}}
\end{equation}
Where $R_P$ is the planet's radius, $D_{cond}$ is the thickness of the conductive layer, $T_S$ and $T_{Top}$ are the surface and top temperature of the convective layer and $k$ is the thermal conductivity coefficient, assumed to be equal to 3.3 W m$^{-1}$K$^{-1}$ \citep{SPOHN2003456}.

The equation for heat transport in the convective, sublayer is:
\begin{equation} \label{eq:convection}
\frac{Q_{conv}}{4 \pi (R_P - D_{cond})^2} = k\frac{(T_m - T_{top} )}{D_{conv}} \left( \frac{Ra}{Ra_c}\right)^\beta
\end{equation}
Where $D_{conv}$ is the thickness of the convective layer, $T_m$ is the melting temperature of the ice I, and $Ra$ and $Ra_c$ are the Rayleigh and the critical Rayleigh number (in the order of $10^3$ \citep{SCHUBERT1979192}). The Rayleigh number is given by Equation \ref{eq:rayleigh}. Convection becomes a significant mechanism for heat transport in the ice shell when $Ra \geq Ra_{c}$. If $Ra \leq Ra_{c}$ we assume that the ice shell is purely conductive (Equation \ref{eq:conduction}).
\begin{equation} \label{eq:rayleigh}
Ra = \frac{\alpha\rho_{ice,I} g_p (T_{m}-T_{top}) D_{conv}^3}{\kappa \eta(T_{int})}
\end{equation}
Where $\alpha$ and $\kappa$ are the ice I thermal expansion coefficient and thermal diffusivity equal to $1.6 \times 10^{-4} K^{-1}$ and $1.47 \times 10^{-6}m^2 s^{-1}$ respectively \citep{sohl_2003}. $g_P$ is the surface gravity of each planet and $\eta(T_{int})$ is the viscosity of the convective layer's mean temperature ($T_{int}$). 

In thermal equilibrium, the heat flux ($Q_{eq}$) that is transferred through the conductive layer is equal to the flux convected through the sublayer:

 \begin{equation} \label{eq:eauilibrium_ice_shell}
 Q_{cond} = Q_{conv} = Q_{eq}
\end{equation}

\begin{figure}[!htb]
    \centering
    \includegraphics[width=0.8\columnwidth]{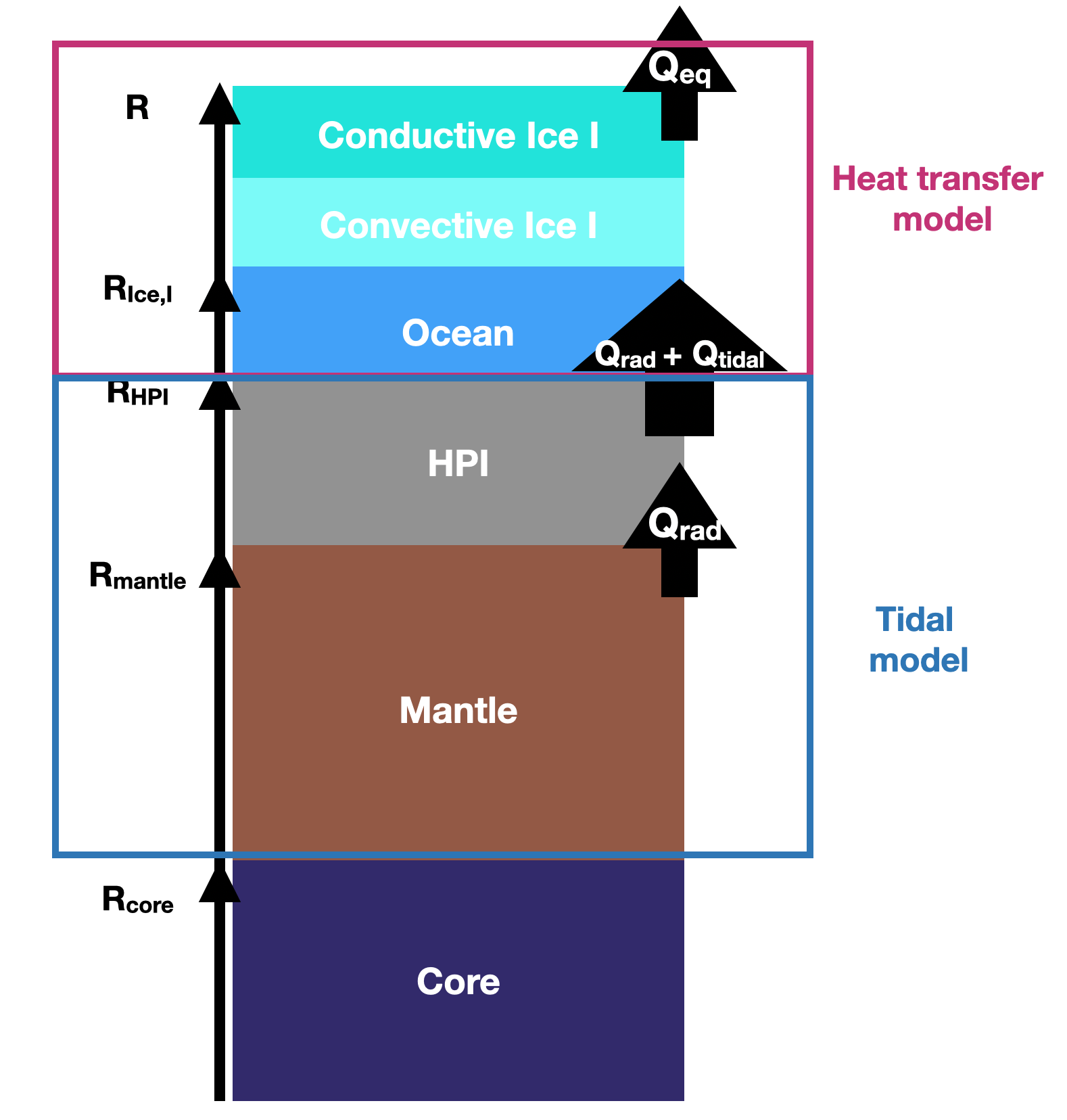}
    \caption{Schematic overview of the interior model of Trappist-1f,g,h. The ice I layer is divided into a conductive and a convective sublayer. The thermal model determines the thickness of the conductive and convective layers, $R_{ice,I}$, $R_{HPI}$, and $Q_{eq}$. The tidal model computes the tidal heating rates in the HPI, and the mantle. The layer boundaries are not to scale. Note that the convective layer is only present for a sufficiently high ($\gtrsim 10^3$) Rayleigh number}
    \label{fig:interiormodel}
\end{figure}

To obtain the thermal equilibrium state of the ice I shell,  we solve Equations \ref{eq:conduction},\ref{eq:convection},\ref{eq:rayleigh} and \ref{eq:eauilibrium_ice_shell} iteratively for the layer thicknesses ($D_{stag}$, $D_{conv}$) and the equilibrium heat flux $Q_{eq}$ for different values of total ice I layer thickness ($D_{ice}$). To do so, we firstly assume a value for $D_{ice}$, which is the free parameter in our interior models. This sets the pressure, and thus, the melting temperature at the bottom of the convective layer ($T_{m}$), at the depth $R_{ice,I}$, as seen in Figure \ref{fig:interiormodel}. The melting temperature is calculated as a function of pressure, as shown in Figure \ref{fig:meltingTvsP} for an ammonia containing ocean of 5 wt.$\%$ \citep{GRASSET2000617}. These concentrations are consistent with those inferred for icy Solar System bodies, like Titan, where ammonia acts as an antifreeze facilitating the existence of subsurface oceans under cold conditions \citep{sohl_2003}. Furthermore, formation beyond the water ice line of Trappist-1f,g,h (see Section \ref{subsection: water_content_formation_scenario}) would promote the accretion of volatile compounds, including ammonia \citep{Oberg2011}. With the assumption that the ocean is isothermal we calculate the pressure at the bottom of the ocean, and thus the thickness of the ocean, and the radius $R_{HPI}$. This process is visualized in the red segment of Figure \ref{fig:flowchart}.
 For more details on the iterative method, we refer the reader to \citet{sohl_2003}. 

\subsection{Tidal response model} \label{tidal_response_model}
The amount of tidal heat dissipated in the interior of a planet is given by Equation \ref{eq:tidalflux} \citep[eg. ][]{Segatz_Spohn_1988}.

\begin{equation}
    \Dot{E} = -\frac{21}{2} \text{Im}(k_2)  \frac{\left(nR\right)^{5}}{G} 
    e^2 
 \label{eq:tidalflux}
\end{equation}   
Where $n$ is the planet's mean motion, $e$ the orbital eccentricity of the planet, and $G$ the universal gravitational constant. We are assuming that the planets are rotating synchronously \citep{Lobo_2023, turbet_2018}. In addition, we assume that the planets have zero obliquity, and therefore have zero obliquity tides.

$\text{Im}(k_2)$ is the imaginary part of the $k_2$ Love number, which describes the  deformation of the planet's gravity field due to the tidal interactions with the host star. $\text{Im}(k_2)$ depends on the interior structure and properties (and hence the mass of the dissipative layers), rheology of the planet and its orbital period \citep[] {Segatz_Spohn_1988, Renaud_2018, Shoji_2014}. Unlike studies that assume a constant $k_2/Q$ (where Q is the tidal quality factor) value for all planets \citep[e.g. ][]{Quick_2023}, in our model ($Im(k_2)=k_2/Q$) is computed self-consistently and varies with the planet’s interior structure, rheology, and orbital period.
We computed the tidal dissipation using the viscolestic theory for self gravitating bodies \citep[see e.g. ][]{peltier_1974, sabadini} with the Maxwell rheological law, which connects stress and strain, and has already been applied to study solar system moons \citep{HUSSMANN2004391, FISCHER199039, Tobie_2019, ROBERTS2008675, rovirra_navaro2022} and exoplanet systems \citep{Rovira-Navarro_2021, Dobos_2015, Hay_2019}. We note that more complex rheological models, such as the Andrade rheological law, have also been used to investigate tidal dissipation in planetary interiors \citep[see e.g. ][]{Kleisioti2023, Henning_2014}. By solving the equations of motion for the deformation of each layer in the Fourier domain via the correspondence principle \citep{peltier_1974}, we obtain the gravitational potential at the surface and thus, the Love number. We use the method described by \citet{Rovira_Navarro_2024} for spherically symmetric bodies, and use the the LOV3D software
package \citep{Rovira_Navarro_2024} to compute the tidal dissipation in each layer for a given planetary structure.

The rheological and physical properties that were used for the thermal and tidal models are presented in Table \ref{table:physical_par}, while the orbital properties of each planet are shown in Table \ref{table:planet_parameters}. We compute tidal dissipation in the mantle, HPI layer, and ice I shell for each interior structure. 
For the equilibrium heat-balance closure in Eq.~\ref{eq: equilibrum_all}, we initially neglect the mantle and ice I contributions and subsequently verify that this approximation is valid over the explored parameter space (Section \ref{tidal_het_results}).
 For the tidal response model we assume an Earth-like mantle and core composition (iron-rich).
The values for mantle viscosity, can range from $10^{15}$ to $10^{22}$Pa s, with $10^{16}$Pa s representing partially molten cases \citep{Barr_2018}. 
A shear modulus of $50$ GPa is commonly adopted for the mantle at temperatures below the solidus \citep{HUSSMANN2002143, Barr_2018}.
Viscosities of the high-pressure ice (HPI) layer, range from $10^{14}$ to $10^{20}$Pa s \citep{Barr_2018}, with $10^{14}$ being close the melting point. The HPI shear modulus was taken to be $\mu_{HPP} \approx 3.5 \times 10^{10}$ Pa, assumed independent of temperature, as in \citet{Barr_2018}. 
Viscosities of Ice I range between $10^{13}$ and $10^{16}$ Pa s, and shear modulus estimates for Ice I typically range from $3.3$ to $4.8$ GPa \citep{HUSSMANN2002143, Barr_2018}.

\begin{table}[htb]
\small
\caption{Average physical and rheological properties of the interior model layers}
\centering
\begin{tabular}{c c c c c c}
\hline\hline
Parameter + Symbol & Value  & Reference \\ [0.5ex] 
\hline
HPI Viscosity ($\eta_{HPI}$) & $10^{14}$ Pa s & \citep{Barr_2018}\\
Mantle Viscosity  ($\eta_{m}$) & $10^{20}$ Pa s &  \citep{lau_2016}\\
Ice I Viscosity ($\eta_{iceI}$) & $10^{14}$ Pa s & \citep{Barr_2018}\\
HPI shear modulus ($mu_{HPI}$) & $3.5 \times 10^{10}$ Pa & \citep{Barr_2018} \\
Mantle shear modulus ($mu_{m}$) & $7 \times 10^{10}$ Pa & \citep{Segatz_Spohn_1988} \\
Ice I shear modulus ($mu_{iceI}$) & $3.3 \times 10^{9}$ Pa& \citep{HUSSMANN2002143}\\
HPI density ($\rho_{HPI}$) &  1310 kg $m^{-3}$ & \citep{sohl_2003}  \\
Mantle density ($\rho_{m}$) &5000 kg $m^{-3}$ & \citep{Barr_2018} \\
Ice I density ($\rho_{ice,1}$) &917 kg $m^{-3}$  & \citep{sohl_2003}\\
Ocean density ($\rho_{ocean}$) &  950 kg $m^{-3}$  & \citep{sohl_2003}\\
Core density ($\rho_{core}$)&12050 kg $m^{-3}$ & \citep{Barr_2018}\\

\hline
\end{tabular}
\label{table:physical_par}
\end{table}

\begin{table*}[t]
\caption{Planet parameters used in the model. }
\centering
\begin{tabular}{c c c c c c}
\hline\hline
Parameter & Symbol & Trappist-1f & Trappist-1g & Trappist-1h \\ [0.5ex] 
\hline
Planet mass [$M_{\oplus}$] & $M_p$ & 1.039 $\pm$ 0.031 & 1.321 $\pm$ 0.038 & 0.326 $\pm$ 0.020\\
Planet radius [$R_{\oplus}$] & $R_p$  & 1.045$^{+0.013}_{-0.012}$  & 1.129$^{+0.015}_{-0.013}$  & 0.775$\pm0.014$  \\
Equilibrium temperature [K] & $T_{eq}$  & 192 & 174 & 151 \\
Orbital period [days] & $P$  & 9.20 & 12.35 & 18.77 \\
Orbital eccentricity & $e$  & 0.01007 & 0.00208 & 0.00567 \\

\hline
\end{tabular}

\caption*{The planet as well as the star parameters that were used in order to calculate the planets' equilibrium temperatures are taken from \citet{Agol_2021} to account for possible surface contamination and the lower reflectivity of ice at near-infrared wavelengths, where M-dwarf spectra peak \citep{shields2013}. For the calculation of $T_{eq}$ we assumed $A = 0.4$. Eccentricity values are taken from \citet{grimm_2018}}
\label{table:planet_parameters}

\end{table*}

\subsection{Interior heat balance and structure calculation} \label{thermal_equilibrium}

The heat transfer model described in Section \ref{sec:heat_tranfer_model} provides the heat flux for a given value of the ice I layer thickness ($D_{ice}$), and the associated thermal equilibrium heat output ($Q_{eq}$). In this manner, $Q_{eq}$ is a function of $D_{ice}$. For the entire planet to exist in equilibrium, $Q_{eq}$ should be equal to the amount of heat that is produced in the interior. Our aim is to determine the mass and thickness of each interior layer - the core ($m_c$), mantle ($m_m$), and high-pressure ice (HPI) layer ($m_{\mathrm{HPI}}$) - for a range of ice shell thicknesses ($D_{\mathrm{ice}}$), such that the total internal heat production is consistent with $Q_{eq}$ and measured planetary mass and radius.

We focus on two dominant sources of interior heat: tidal heating in the HPI layer and radiogenic heating in the silicate mantle, and assume that the heat flux through the ice shell ($Q_{\mathrm{eq}}$) must equal the sum of the radioactive decay ($Q_{rad}$) that is produced by the accumulation of radioactive elements in the silicate mantle, and the tidal heat that is produced in the high pressure ice layer ($Q_{tidal,HPI}$). Equation \ref{eq: equilibrum_all} is used as a closure condition to identify interior structures in thermal equilibrium. We then verify its validity by recomputing tidal dissipation in all layers (mantle, HPI, and ice I) for the derived structures and by exploring a wide range of rheological parameters via a Monte Carlo sensitivity analysis (Section \ref{sensitivity_analysis}).
\begin{equation}\label{eq: equilibrum_all}
Q_{\mathrm{eq}} = Q_{\mathrm{tidal,HPI}} + Q_{\mathrm{rad}} 
\end{equation}

For the calculation of $Q_{rad}$, we follow the approach of \citet{schubert1986}, and use Equation (1) from \citet{HUSSMANN2004391}, which expresses the radiogenic heat production as a function of mantle mass ($m_m$). We assume a mantle with chondritic composition and a system age of 7.6 Gyr \citep{Burgasser_2017}. We calculate $Q_{\mathrm{tid,HPI}}$ using Equation~\ref{eq:tidalflux}, which provides a direct relationship between $Q_{\mathrm{tid,HPI}}$ and the mass of the high-pressure ice layer ($m_{\mathrm{HPI}}$). We note that additional mechanisms, such as hemispheric tectonic driven by strong day-night temperature contrasts on exoplanets, may also influence heat transport \citep{Meier_2021}. 

To constrain the full interior structure, we can apply the following constraints, which describe the mass conservation (Equation \ref{eq: mass_conservation}), and the geometry of the system (Equation \ref{eq:geometry}), while assuming homogeneous layers with constant density. 

\begin{equation}
 M_{p} = m_c + m_{m} + m_{ocean} + m_{ice,I} + m_{HPI} \label{eq: mass_conservation}
\end{equation}

 

We solve for $R_{HPI}$, defined as the radius of the boundary between the high-pressure ice layer and the ocean, using the volumes of the layers interior to this interface:

 \begin{equation} \label{eq:geometry}
R_{HPI}  = \left( \frac{3}{4\pi} \right)^{\frac{1}{3}} \left( \frac{m_{HPI}}{\rho_{HPI}}  +  \frac{m_{m}}{\rho_{m}}  + \frac{m_{c}}{\rho_{core}} \right)^{\frac{1}{3}}
\end{equation}
Where $m_{ice,I}$ and $m_{ocean}$ is the mass of the ice I shell, and the subsurface ocean. The rest of parameters in Equations \ref{eq: mass_conservation}, \ref{eq:geometry} are described in Table \ref{table:physical_par}. 

\begin{figure}[!htb]
    \centering
    \includegraphics[width=0.8\columnwidth]{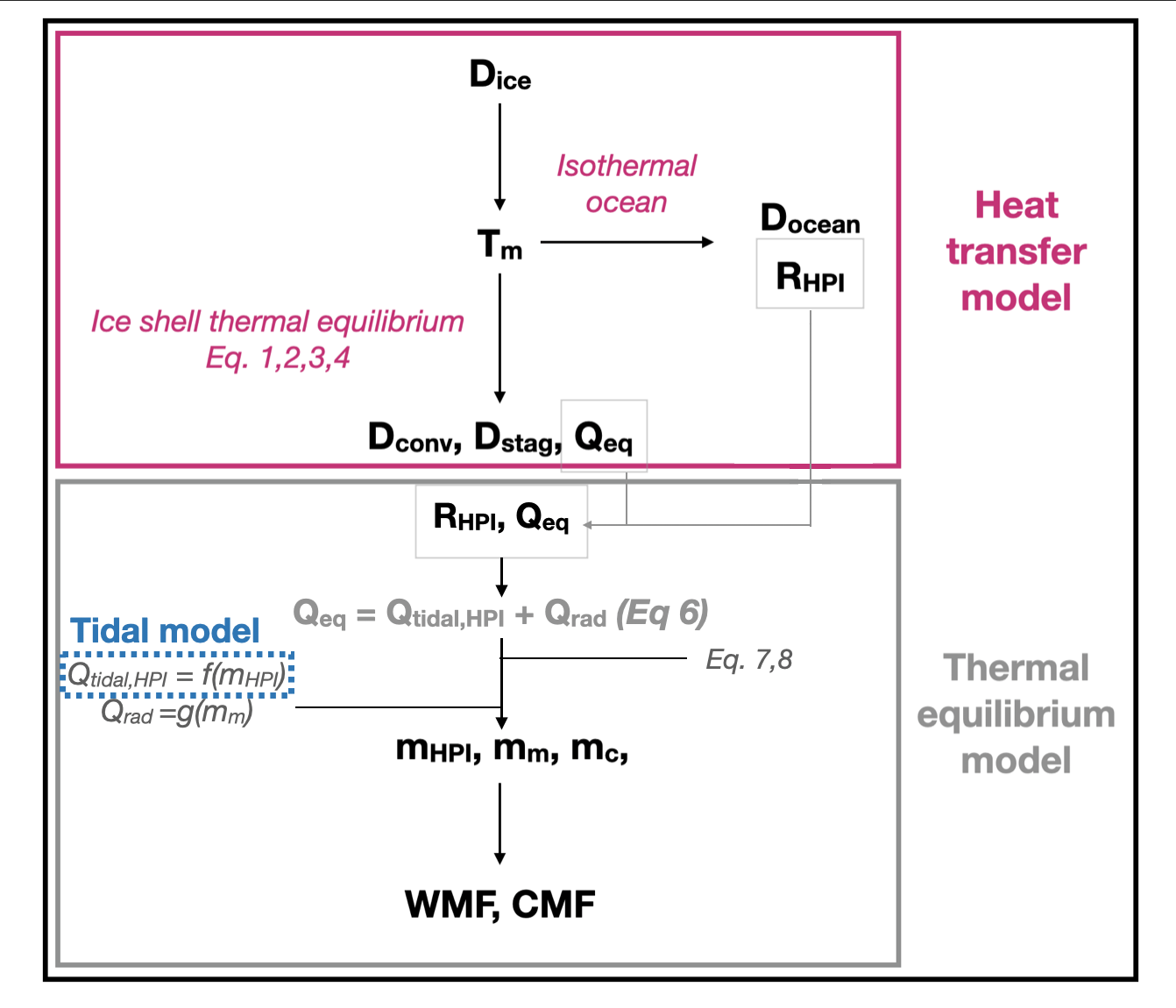}
    \caption{Schematic of the parameter calculation process. The red segment visualizes the heat transfer model in the outer ice shell, which is described in Section \ref{sec:heat_tranfer_model}. The grey segment describes the process of calculating the thermal equilibrium in the entire planet (Section \ref{thermal_equilibrium}). The latter thermal equilibrium dictates that $Q_{eq} = Q_{tidal,HPI} + Q_{rad}$}
    \label{fig:flowchart}
\end{figure}

Since $Q_{tid,HPI}$ is a function of $m_{HPI}$ (see Section \ref{tidal_response_model}) and $Q_{rad}$ is a function of $m_{m}$, we solve the system of equations \ref{eq: equilibrum_all}, \ref{eq: mass_conservation}, \ref{eq:geometry} for $m_m$, $m_{HPI}$, and $m_c$ for a range of $D_{ice}$. Thus, we calculate $m_c$, $m_m$, $m_{HPI}$, $Q_{rad}$, and $Q_{tidal,HPI}$ as a function of $D_{ice}$. This allows us to derive the Core Mass Fractions (CMF) and Water Mass Fractions (WMF) values as a function of $D_{ice}$. The gray boxed region of Figure \ref{fig:flowchart} visualises the calculation process described in this Section. 

\begin{figure}[!htb]
    \centering
    \includegraphics[width=0.98\columnwidth]{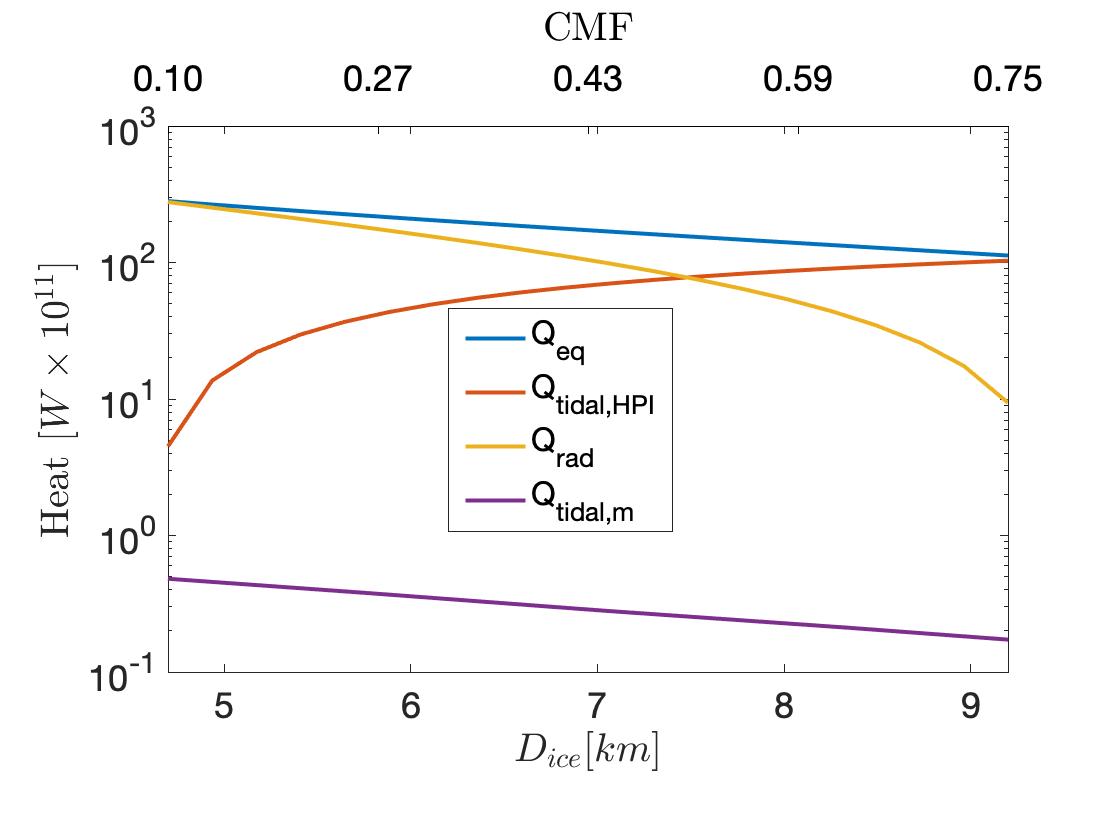}
    \caption{Comparison of heat contribution from several sources for Trappist-1f as a function of ice shell thickness. Tidal heat in the mantle is negligible in the entire parameter space, while radiogenic heat and tidal heat in the ice shell both contribute significantly to the ice sheet equilbrium heat. Trappist-1f due to its size and eccentricity has the highest tidal heat of the three planets.}
    \label{fig:heatcomparison}
\end{figure}

We finally assess whether the assumption made in Equation \ref{eq: equilibrum_all} is valid. We recompute the expected tidal heating rates in the different layers, as seen in Figure \ref{fig:heatcomparison} 
for the computed interior structures with varying HPI, mantle, and core masses.
In all cases, we conclude that the tidal heating rate in the mantle ($Q_{tidal,m}$) is at least one order of magnitude lower than both the tidal heating in the HPI and the expected radiogenic heat (see Figure \ref{fig:heatcomparison}) , while the tidal heat in the ice I shell ($Q_{tidal,iceI}$) is typically four orders of magnitude smaller than the total heat generated in the interior (see Figure \ref{fig:Tidalheatice1}), validating our initial assumption of neglecting $Q_{tidal,m}$ and $Q_{tidal,iceI}$ in Equation \ref{eq: equilibrum_all}.


\subsection{Sensitivity analysis} \label{sensitivity_analysis}
To evaluate the robustness of our tidal dissipation results and address uncertainties in interior rheology and internal heating, we performed a Monte Carlo sensitivity analysis over a broad range of model parameters. Layer viscosities were varied independently using log-uniform distributions spanning several orders of magnitude. Mantle viscosities were sampled over the range $\log_{10}(\eta_{\mathrm{m}}/\mathrm{Pa\,s}) = 18$-21, HPI viscosities over $\log_{10}(\eta_{\mathrm{HPI}}/\mathrm{Pa\,s}) = 14$-18, and ice~I shell viscosities over $\log_{10}(\eta_{\mathrm{ice\,I}}/\mathrm{Pa\,s}) = 13$-16. These ranges represent values commonly adopted in previous studies of icy bodies \citep{Barr_2018, HUSSMANN2002143}.
In addition, the radiogenic heat flux was varied between half and twice the nominal value used in the reference model, corresponding to
$H = (0.25\text{-}1.0)\times10^{-12}\,\mathrm{W\,m^{-2}}$.
For each realization, we computed the tidal dissipation rates in the mantle, HPI layer, and ice~I shell, as well as the total tidal dissipation $Q_{tidal}$ and the equilibrium heat flux $Q_{\mathrm{eq}}$.

\section{Tidal dissipation and interior constraints for Trappist-1f,g,h} \label{interior_results}
We first demonstrate that neglecting tidal dissipation in the silicate mantle and the ice I shell does not significantly affect the equilibrium heat balance over the parameter space considered here, by quantifying how tidal dissipation is partitioned among the planet’s interior layers. We then assess which layers dominate the tidal energy budget and present the interior structures that can exist in thermal equilibrium.

\subsection{Sensitivity analysis results} \label{tidal_het_results}

\begin{figure}[!htb]
    \centering
        \includegraphics[width=\columnwidth]{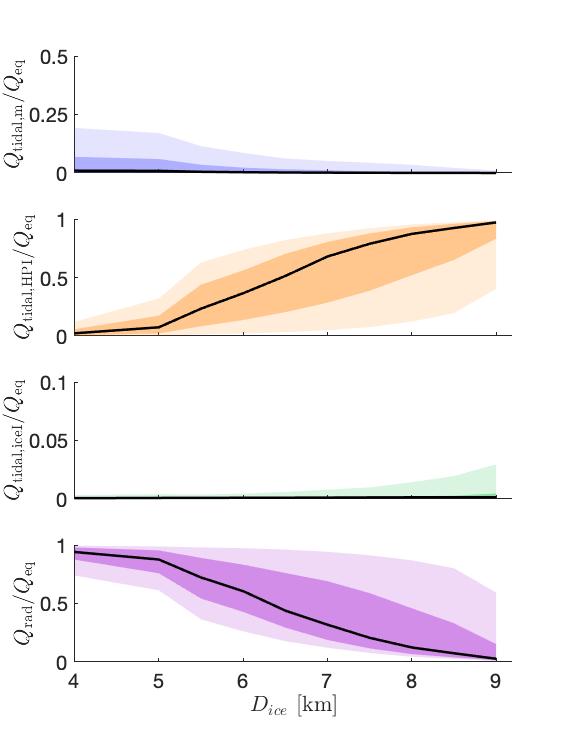}
        \caption{Fractions of $\frac{Q_{\mathrm{tidal}}}{Q_{eq}}$ of the mantle, high-pressure ice layer, and outer ice~I shell, as a function of $D_{\mathrm{ice}}$. The lighter and darker shaded regions indicate the 75\% and 95\% uncertainty intervals derived from the Monte Carlo distribution, respectively.}
        \label{fig:montecarlo}
\end{figure}

\begin{figure*}

    \begin{minipage}{1\textwidth}
        \centering
        \includegraphics[width=\linewidth]{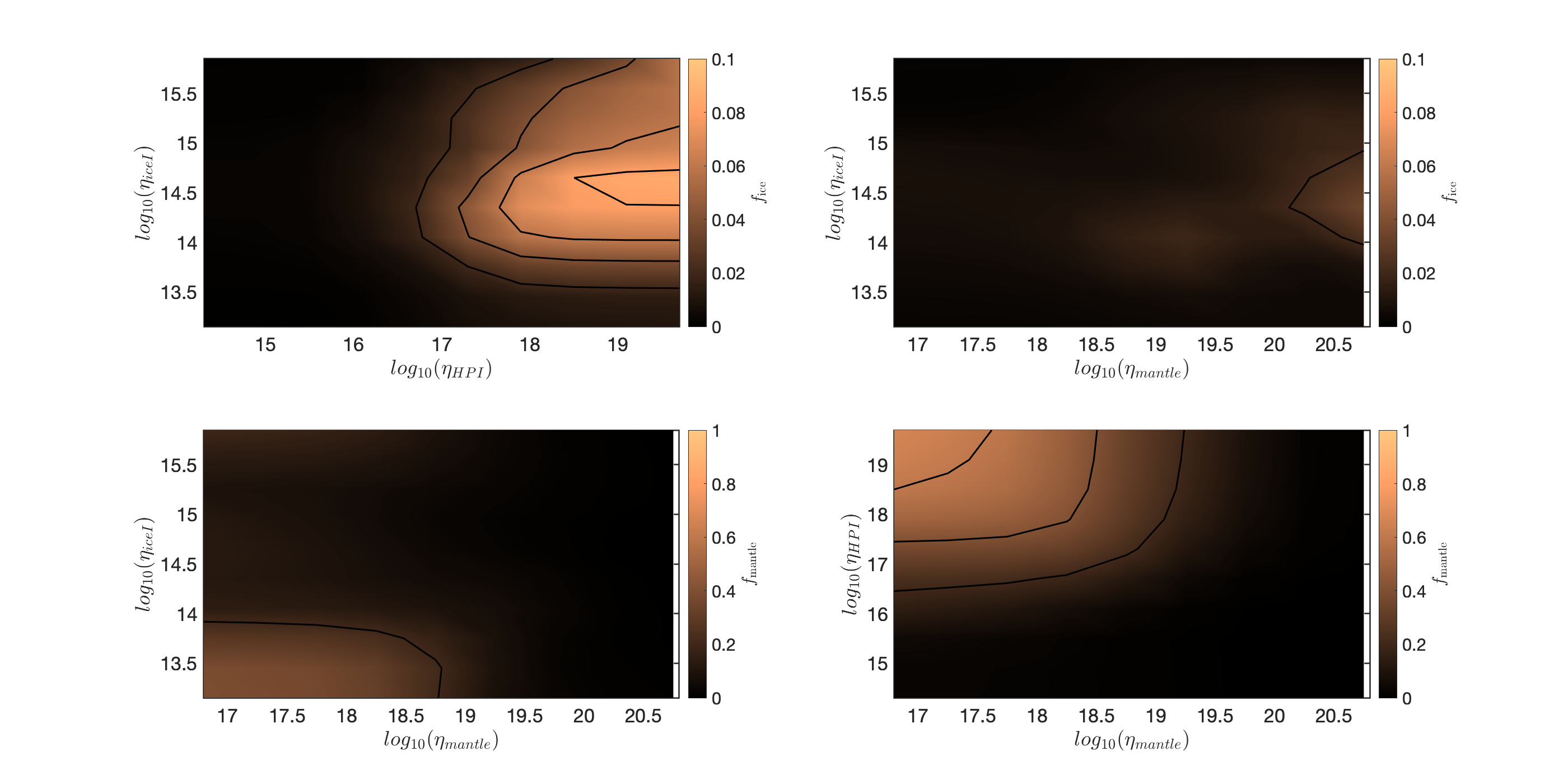}
        \caption{Top: $f_{ice}$ as a function of different layer viscosity. Bottom: $f_{mantle}$ as a function of different layer viscosity. Enhanced tidal dissipation occurs in the ice I shell for stiffer HPI layers. Tidal dissipation in the mantle contributes more to the heat budget for stiffer HPI layers and lower mantle viscosity.}
        \label{fig:n_f_montecarlo}
    \end{minipage}
\end{figure*}

Figure \ref{fig:montecarlo} shows the fractional contributions of tidal dissipation in the mantle, high-pressure ice layer, and outer ice~I shell, normalized by the total tidal dissipation, as a function of ice shell thickness $D_{\mathrm{ice}}$. Across the entire parameter space explored, tidal dissipation within the ice I shell accounts for only a small fraction of the global energy budget. In all realizations, the maximum fraction of tidal dissipation occurring in the ice I shell is
$f_{ice}$ = $Q_{\mathrm{tidal,ice\,I}}/Q_{\mathrm{tidal}} \lesssim 0.1$.
Because radiogenic heating is a significant fraction of the total energy budget, the relative contribution of ice I tidal dissipation to the total heat budget is even smaller, yielding
$Q_{\mathrm{tidal,ice\,I}}/Q_{\mathrm{total}} \lesssim 0.04$. The largest fractional contributions from the ice I shell occur primarily for models with relatively stiff HPI layers (see Figure \ref{fig:n_f_montecarlo} which illustrates the dependence of tidal dissipation partitioning on layer viscosities). In these cases, reduced deformation in the HPI layer allows a greater fraction of the tidal strain to be accommodated within the overlying ice~I shell. Nevertheless, even under these conditions, tidal dissipation in the ice I shell is a small portion of the total heat budget.

Enhanced tidal dissipation in the mantle occurs only in a subset of models with small core mass fractions (CMF $\lesssim 0.1$) and relatively low mantle viscosities. In these cases, the mantle contribution can reach
$Q_{\mathrm{tidal,m}}/Q_{\mathrm{eq}} \sim 0.2$, since radiogenic heating in the mantle also contributes significantly to the total heat budget. Such conditions are mostly achieved for stiff HPI layers combined with smaller mantle viscosities (Figure \ref{fig:n_f_montecarlo}). If partial melting were to occur in the mantle, the fraction of tidal dissipation localized there would likely increase further.

Overall, these results demonstrate that, for the interior structures considered here, tidal dissipation is not dominated by the ice I shell, in contrast to inferences drawn from Solar System satellites \citep[see e.g. ][]{sohl_2003}. This difference arises primarily from the specific interior configurations explored in this study, in particular the presence of very thick high-pressure ice (HPI) layers, in some cases exceeding 
$\sim$2000 km in thickness. Such a thick viscoelastic HPI layer represents a large volume of deformable solid and can therefore contribute substantially to the total tidal dissipation budget, reducing the relative importance of the overlying ice I shell.


\subsection{Possible interior structures of the planets Trappist 1 - f,g,h}\label{sec:results3.1}
\begin{figure}[!htb]
    \centering
    \includegraphics[width=0.99\columnwidth]{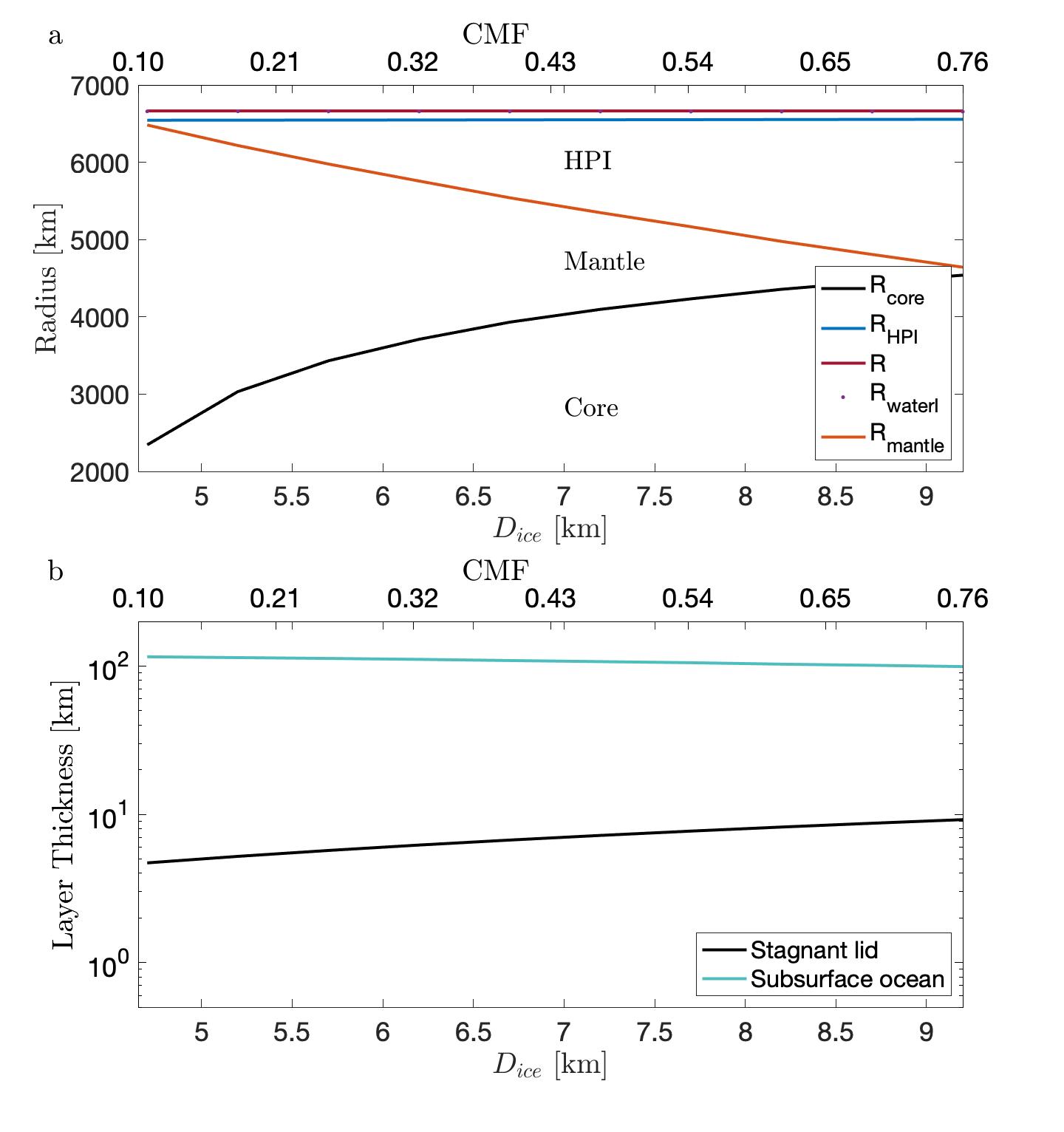}
    \caption{Interior structure as a function of $D_{ice}$ for the planet Trappist-1f. a. Radii of different layers as a function of $D_{ice}$. b. Thickness of the stagnand lid, convective and ocean layers as a function of $D_{ice}$.} 
    \label{fig:interiorf}
\end{figure}

In Figure \ref{fig:interiorf}a, we present the interior structure of Trappist-1f ($R_{core}$, $R_{mantle}$, $R_{HPI}$) as a function of $D_{ice}$ (see Section \ref{thermal_equilibrium}).
Figures \ref{fig:interiorg}, \ref{fig:interiorh} present the corresponding Figures for Trappist-1,g and h.
A given $D_{ice}$ thickness can transport an amount of heat equal to the radiogenic heat generated in the mantle and the tidal heat produced in the HPI layer.
Thicker ice I layers can transport smaller amounts of heat. The latter translates to  a structure with higher $\frac{M_{HPI}}{M_{Mantle}}$ ratios, as seen in Figure \ref{fig:heatcomparison}.
Thus, as $D_{ice}$ increases, the ratio $\frac{M_{HPI}}{M_{Mantle}}$ also increases (Figure \ref{fig:interiorf}).
The various interior structures are consistent with the observed densities of the planets.
As a result, the calculated range for ice shell thickness is limited; thicker ice shells reach thermal equilibrium at lower $Q_{eq}$ values.
Achieving such equilibrium requires a larger core, since a smaller mantle reduces radiogenic heating.
This, in turn, prevents the total internal heat, radiogenic from the mantle and tidal from the high-pressure ice (HPI) layer, from exceeding the heat transfer capacity of the ice shell.
Thus, the high-pressure ice (HPI) layer expands rapidly with $D_{ice}$, compensating for the increased core size.

The right horizontal-axis limit of Figure \ref{fig:interiorf} corresponds to the case of a planet without a mantle layer, meaning that the iron core would be in direct contact with the HPI (CMF = 0.75). 
The left horizontal-axis limit is consistent with an interior structure that lacks a HPI layer. Such an interior structure corresponds to the minimum WMF. Smaller WMF's would not be compatible with Trappist-1f,g,h having an icy shell with a subsurface liquid ocean under the assumptions of our model. In particular, changing the assumptions for the physical and rheological parameters in Fig. \ref{table:physical_par} will change the specific numerical values we obtain, but the trends and overall magnitude of the quantities we present will be representative for a body with the interior layer structure as defined in our model. Without an HPI layer, the subsurface ocean would directly interact with the mantle. This interaction is crucial for astrobiology and habitability studies, as it allows essential elements for the development of life to potentially enrich the subsurface ocean, similar to what is believed to occur on Europa \citep{Hand_2022}. Recently \citep{Hernandez2022} suggested that even when high-pressure ice layers are present, these do not fully block chemical communication. This suggests that interior-ocean exchange of electrolytes and essential elements may remain possible across a wide range of planetary conditions.

Figure \ref{fig:interiorf}b shows the thickness of the stagnant lid, convective layer, and ocean layer, as a function of $D_{ice}$. If Trappist-1f has a CMF of 0.5 and an ice shell in thermal equilibrium, the ocean, approximately 100 km thick, would lie about 6 km beneath an almost entirely conductive ice shell. The extremely thin convective shell thickness
supports the assumption of a conducting ice I layer made in this work (Section \ref{tidal_response_model}), in contrast to the convective, thicker ice shells observed on icy moons within the solar system \citep[see e.g. ][]{HUSSMANN2002143}.

We present the thermal equilibrium interior structure parameters for Trappist-1f,g,h in Tables \ref{table:RESULTSf}, \ref{table:RESULTSg}, \ref{table:RESULTSh} respectively for CMF=0.18, 0.32, and 0.5 assuming that the planets have not been
exposed to events that could have stripped away their mantle, such as mantle evaporation \citep{CAMERON1985285}. For the planets Trappist-1f and g, even in the case of a large CMF equal to 0.5, a subsurface ocean would exist below an ice shell of thickness $\lesssim$ 10\,km. Such a thin ice I shell hints that the ocean is more likely to be in direct contact with the surface through cryovolcanism. As a result, a plume  would likely be composed of material originating directly from the subsurface ocean, similarly to the cryovolcanic mechanism on Enceladus \citep{Quick_2023, Porco_2014}. Trappist-1h, the least massive of the three planets, exhibits equilibrium interior structures with an ice layer thickness $\gtrsim 10$ km. This raises the possibility that any cryovolcanism on Trappist-1h could resemble that of Europa, coming from water-pockets in the ice \citep{Quick_2023}. The ice shell I thicknesses that we obtain are mostly in agreement with the results of \citep{Quick_2023}. 
For some interior configurations, particularly those with thicker Ice I shells, our modeled tidal heating rates for Trappist-1f ($\approx$ 10 TW) fall within the same order of magnitude as previous estimates by \citet{Bolmont2026}, who constrain tidal heating in Trappist-1f between $\approx$ 10-30 TW.

In the following section we discuss whether cryovolcanism could be observed with spectroscopic observations on Trappist-1f,g,h. 

\begin{table*}[t]
\caption{Interior and thermal parameters of Trappist-1f in thermal equilibrium for different values of CMF.}
\centering
\begin{tabular}{c c c c c c}
\hline\hline
Parameter & Symbol  & CMF = 0.11 (min)& CMF = 0.18 &CMF = 0.32& CMF = 0.50 \\ [0.5ex] 
\hline
Thickness of ice I layer [km] & $D_{ice,I}$ & 4.7 & 4.9 & 5.6 & 6.8 \\
Ocean thickness [km] & $D_{ocean}$  &117.4 & 117.1 & 115.4 & 111.5 \\
Water mass fraction & WMF & 0.019 & 0.038 & 0.085  & 0.14 \\
\hline
\hline
\end{tabular}
\label{table:RESULTSf}
\end{table*}

\begin{table*}[t]
\caption{Interior and thermal parameters of Trappist-1g in thermal equilibrium for different values of CMF}
\centering
\begin{tabular}{c c c c c c}
\hline\hline
Parameter & Symbol & CMF = 0.09 (min) &CMF = 0.18 &CMF = 0.32& CMF = 0.50 \\ [0.5ex] 
\hline
Thickness of ice I layer [km] &  $D_{ice,I}$ &5.2 &5.9 & 7.2 & 9.2 \\
Ocean thickness [km] & $D_{ocean}$ & 112.5 &110.0 & 104.4 & 95.6 \\

Water mass fraction & WMF &0.011 & 0.030 & 0.059 & 0.080 \\
\hline
\hline
\end{tabular}
\label{table:RESULTSg}
\end{table*}

\begin{table*}[t]
\caption{Interior and thermal parameters of Trappist-1h in thermal equilibrium for different values of CMF}
\centering
\begin{tabular}{c c c c c c c}
\hline\hline
Parameter & Symbol &CMF = 0.02 (min) & CMF = 0.18 &CMF = 0.32& CMF = 0.50& \\ [0.5ex] 
\hline
Thickness of ice I layer [km] & $D_{ice,I}$ & 9.4 &11.5 & 13.7 & 18.0 \\
Ocean thickness [km] & $D_{ocean}$ & 111.0&108.4 & 103.5& 92.0 \\

Water mass fraction & WMF & 0.080& 0.095 & 0.118 & 0.150 \\
\hline

\hline
\end{tabular}
\label{table:RESULTSh}
\end{table*}

\section{Modelling observations of cryovolcanic activity} \label{sec:cryovolcanism_methods}

Cryovolcanic activity has been observed on Saturn's moon, Enceladus \citep{porco_helfenstein2006, spencer2018}, and is also suspected on Jupiter's moon, Europa, based on multiple detections \citep{Roth2014, Sparks_2017, papagini2020}. In the case of Enceladus, a subsurface ocean
is located $\lesssim$ 10 km beneath the south pole \citep{iess2014}. 
The thin ice shell of Enceladus allows cracks to propagate fully downward to the ocean due to the long-term evolution of the ocean pressure and ice shell thickness \citep{Rudolph2022}. 
Cryovolcanic activity on Enceladus is thought to be continuous, while its intensity appears to be correlated with the orbital period \citep{INGERSOLL2020} and tidal stresses \citep{hedman2013observed, soucek2024variations}. On the other hand, cryovolcanic eruptions on Europa are sporadic \citep{Sparks_2017}. In particular, \citep{papagini2020, Phillips2000} show that large cryovolcanic eruptions are likely rare, and \citet{FAGENTS2000} suggest that smaller eruptions are likely to occur more frequently. Given an ice shell of $\approx$ 25–30 km \citep{QUICK201516}, cryovolcanic activity on Europa is thought to originate from water pockets that are located within the ice shell \citep{Fagents2003,manga_2007}.

Similar geological features, including volcanism and cryovolcanism, may be present on exoplanet systems \citep{Kleisioti2024, Quick_2023, Quick_2020, Oza_2019, zuWestram2024, guenther2019}. However, predicting cryovolcanic outgassing on cold ocean exoplanets is challenging because it can be influenced by poorly constrained parameters and interior processes, including the total internal heat production, possibly the fraction of that heat generated by tidal dissipation, and the thickness of the outer ice shell. 
 While surface fractures, like Enceladus' tiger stripes, may form solely due to tidal deformation of the ice shell, radiogenic heating can also contribute to maintaining subsurface liquid reservoirs \citep{Leone2021}. 
If tidal dissipation is sufficient to create cracks, radiogenic heat may further sustain or enhance internal melting, and together these heat sources can maintain subsurface oceans that enable cryovolcanic activity \citep{INGERSOLL2020, Neveu2019}.

The total volcanic outgassing $M_{volc}$ that is driven by $\dot{E}_{Total}$ can be expressed in terms of the gravitational binding energy of a water molecule to the planet ($U$):

\begin{equation}\label{eq:mvolc_total2}
\dot{M}_{Volc}  \approx  \eta 
 m_{H_20} 
 \frac{\dot{E}_{Total}}{U}%
\end{equation}

Where  $\eta$ is a factor describing the efficiency that the internal energy is used to power volcanic outgassing \citep{Oza_2019}, $m_{H_20}$ is the mass of one water molecule. The gravitational binding energy $U$ is equal to $\frac{G M_p m_{H_20}}{R_p}$ \citep{johnson2015}. An efficiency factor of $\eta = 1$ would represent the theoretical upper limit, where all the available energy is used for the outgassing of water. In reality, this value is much lower. Using a mass loss rate of 300 kg s$^{-1}$ \citep{villanueva2023jwst} for Enceladus and an internal power of 35 GW \citep{HEMINGWAY2019}, one obtains an efficiency factor for Enceladus on the order of $10^{-3}$.

Given their relatively close-in orbits and the active nature of TRAPPIST-1, exospheres may form not only through outgassing from cryovolcanic activity, but also via the mass loss through thermal escape caused by XUV activity \citep{Wordsworth_2018, Luger_2015}(see also Section \ref{subsection: water_content_formation_scenario}.  
 \citet{Quick_2023} modeled this latter phenomenon by calculating the water mass loss rate under the XUV flux of TRAPPIST-1, and translating it into column densities produced by the XUV-driven escape.

The total H$_2$O column density is computed assuming a steady-state balance between
continuous surface and interior supply and XUV-driven loss processes. Specifically,
the observable column density is given by

\begin{equation}
\label{eq:Ntot_dayside}
N_{\rm total}
=
\frac{\dot{M}_{\rm volc}\,\tau_{\rm XUV}}{4\pi R_p^2\,m_{\rm H_2O}}
\;+\;
\frac{\dot{M}_{\rm sub}\,\tau_{\rm XUV}}{2\pi R_p^2\,m_{\rm H_2O}} .
\end{equation}

where $\dot{M}_{\mathrm{volc}}$ and $\dot{M}_{\mathrm{sub}}$ are the mass
rates of water vapor due to cryovolcanic outgassing and thermal sublimation,
respectively. In Eq. \ref{eq:Ntot_dayside} volcanic outgassing is globally distributed while surface sublimation occurs only over the illuminated hemisphere. The quantity $\tau_{\mathrm{XUV}}$ represents the effective
molecular lifetime against XUV-driven photochemical loss.

We parameterize the sublimation flux, assuming it occurs primarily on the illuminated hemisphere and using the analytical form
\begin{equation}
\dot{M}_{sub} = 2 \pi a_{sub} T_{eq}^{1/2} \exp\!\left(-\frac{b}{T_{eq}}\right),
\end{equation}
where $a_{sub} = 1.9 \times 10^{32} \text{cm}^{-2} \text{s}^{-1}$ and $b = 8500 \text{K}$ are empirically chosen coefficients to represent a low-sublimation case as was simulated for H$_2$O at Ganymede's leading hemisphere   \citep{LEBLANC2017, oza2017}. A second more realistic case, for pure water ice sublimation following \citet{bob81} considers a canonical $b = 6146 \text{K}$ for the same $a_{sub}$. This substantially increases the expected column density of water vapor, allowing for a steady-state atmosphere to accumulate. Deviations from this column density in future observations, are therefore constraints on the purity of ice on the surfaces of icy exoplanets in principle. We test both sublimation cases and consider our low-sublimation results as representing a conservative, lower bound on the expected atmospheric column density from pure water ice sublimation.

Following \citet{Quick_2023}, we
approximate this lifetime using the H$_2$O photoionization timescale scaled from Europa,  as described in Equation \ref{eq:tau_xuv}.
\begin{equation}
\tau_{\mathrm{XUV}} 
= 4.54\times 10^{7}\,\mathrm{s}\,
\left(\frac{a}{a_{\mathrm{Europa}}}\right)^{2},
\label{eq:tau_xuv}
\end{equation}
where $a$ is the planet's semimajor axis and $a_{\mathrm{Europa}}$ is Europa's
orbital distance. This scaling assumes that the relevant ionizing flux varies as
$a^{-2}$ scaled to Europa's reference value \citep{SHEMATOVICH2005}. Future works considering the \textit{bare rock} or \textit{bare ice} scenarios of airless exoplanets and exomoons should consider the detailed electron-impact dissociation, and plasma destruction rates of water as is estimated in 3-D exosphere models for Europa \citep{oza2019europa}.

We consider three limiting atmospheric configurations for a fixed volatile inventory $n_{tot}$: (1)hydrostatic, representing the maximum possible collisional redistribution, (2)globally uniform sputtered exospheres, (3) and spatially localized sputtered exospheres, to explore how different distributions of cryovolcanically produced water affect observable transmission spectra. These scenarios span the range of plausible density structures that may arise from cryovolcanic activity and allow us to assess how different distributions of cryovolcanic outgassed material influence detectability.

\subsection{Hydrostatic atmospheres} \label{sec:hydrostatic}
We first examine the hydrostatic limit of an atmosphere composed exclusively of water vapor. Under this assumption, the total surface pressure reduces to the partial pressure of water,
\begin{equation}
    P_{\mathrm{total}}  = N_{\mathrm{total}}\, m_{\mathrm{H_2O}}\, g_P,
\end{equation}
where $g_P$ is the planetary surface gravity.

Assuming an isothermal atmosphere in hydrostatic equilibrium, and an exponential decay model for the density of the planet's atmosphere, we get a density profile of the outgasses water. We set the atmospheric temperature equal to the planetary equilibrium temperature $T_{\mathrm{eq}}$ (Table~\ref{table:planet_parameters}). This hydrostatic density structure is used to compute transmission spectra with the radiative transfer code \textit{petitRADTRANS} \citep{prt2019, prt2020, prt2022}, employing the POKAZATEL linelist \citep{Polyansky2018} for water opacities. Spectra are modeled over the wavelength range 0.2--2.5\,$\mu$m, covering the dominant H$_2$O absorption features, and are subsequently forwarded to the \textit{PandExo} exposure time calculator \citep{pandexo} to generate mock JWST datapoints using the JWST/NIRISS Substrip 256 instrument.

\subsection{Globally uniform sputtered exospheres}
We next consider the case in which the water molecules are uniformly distributed across the planetary surface, producing a globally homogeneous collisionless exosphere.
The quantity $N_{\mathrm{volc}}$ represents the vertical column density of H$_2$O, i.e. the number of water molecules per unit surface area. To obtain the total number of outgassed water molecules, $n_{\mathrm{tot}}$, we assume that the volatile material is uniformly distributed over the planetary surface. In this case, the relevant area is the full spherical surface, $A = 4\pi R_p^2$, and the total number of molecules is
\begin{equation}
    n_{\mathrm{tot}} = 4\pi R_p^2 \, N_{\mathrm{total}}.
\end{equation}

This $n_{\mathrm{tot}}$ is then used as input to the  radiative transfer tool \textsc{Prometheus}  \citep{Gebek2020} to generate synthetic transmission spectra of a water exosphere as follows.

\textsc{Prometheus} computes transmission spectra for gaseous media in arbitrary geometries, assuming line-of-sight optical depth integration over a user-specified 3D density distribution. Because a cryovolcanic plume expands collisionlessly above the exobase and rapidly becomes non-hydrostatic, we treat the outgassed water as an extended exosphere rather than as an isothermal hydrostatic atmosphere. The density distribution is therefore modeled using a power-law scenario, which assumes a spherically symmetric exosphere of the form
\begin{equation}
    n(r) = n_{0} \left( \frac{R_{p}}{r} \right)^{q},
\end{equation}
where $q$ is the power-law index, r the radial distance from the planet's center, and $n_{0}$ is determined by the total number of absorbing molecules in the exosphere.
We adopt a power-law index $q = 6$ \citep{Gebek2020}, which controls how steeply the exosphere density falls
with radius, and a pseudo-temperature of 200\,K for the H$_2$O absorber, which determines the
thermal Doppler width of individual lines. Steep density profiles with indices $\approx 6$ are indicative of strong ion–neutral scattering, where neutral molecules interact efficiently with the surrounding plasma and are removed on short timescales \citep{Johnson1990}. We compute the transmission spectra for the globally uniform sputtered exopsheres using the same opacity treatment, wavelength coverage, and JWST instrument simulation setup as the one adopted for the hydrostatic atmospheres in Section \ref{sec:hydrostatic}.

\subsection{Effect of spatial localization}
In the preceding analysis, we assumed that material outgassed via cryovolcanic activity is uniformly distributed over the planetary surface, corresponding to a globally averaged column density. This assumption provides a conservative lower limit on detectability. However, cryovolcanic activity are expected to be spatially structured, producing localized denser areas analogous to plume-like configurations.
To quantify the effect of such localization, we introduce a surface filling factor $f$, defined as the fraction of the planetary surface area $4\pi R_p^2$ over which the sputtered material is distributed. For a fixed  number of cryovolcanically produced molecules ($n_{\mathrm{volc}}$), the effective column density associated with a localized cloud is then given by
\begin{equation}
    N_{\mathrm{volc,\,loc}} = \frac{n_{\mathrm{volc}}}{4\pi R_p^2\,f} 
\end{equation}
such that $f=1$ corresponds to a globally uniform cloud, while $f\ll1$ represents increasingly localized, plume-like distributions.
The total effective column density becomes

\begin{equation}
\label{eq:Ntot_loc}
N_{\rm total,loc}
=
\frac{\dot{M}_{\rm volc}\,\tau_{\rm XUV}}{4\pi R_p^2\,m_{\rm H_2O}}
\frac{1}{f}
\;+\;
\frac{\dot{M}_{\rm sub}\,\tau_{\rm XUV}}{2\pi R_p^2\,m_{\rm H_2O}} .
\end{equation}

Concentrating a fixed inventory into a smaller surface area increases the line-of-sight column density and therefore enhances the observability of spectral features. When $f = 1$, the expression becomes equivalent to Equation~\ref{eq:Ntot_dayside}, corresponding to the globally uniform redistribution case. Throughout this work, this limiting case is adopted as a conservative baseline.  Physically plausible values of $f$ are uncertain, since they depend on the number, geometry, and temporal variability of active vents. We adopt a value of $f=0.004$ as a limiting case of a localized configuration corresponding approximately to the fractional surface area associated with an individual tiger-stripe source region \citep{porco_helfenstein2006}. More generally, values of $f\sim10^{-2}$-$10^{-1}$ may represent broad regional localization, whereas $f\lesssim10^{-3}$ would correspond to highly localized vent-like sources.

\section{Transmission spectroscopy signatures of cryovolcanic outgassing} \label{sec:transmission_spectra}

\begin{figure}[!htb]
    \centering
        \begin{subfigure}{0.99\columnwidth}
        \centering
        \includegraphics[width=\linewidth, trim=0cm 0cm 0cm 11.7cm, clip]{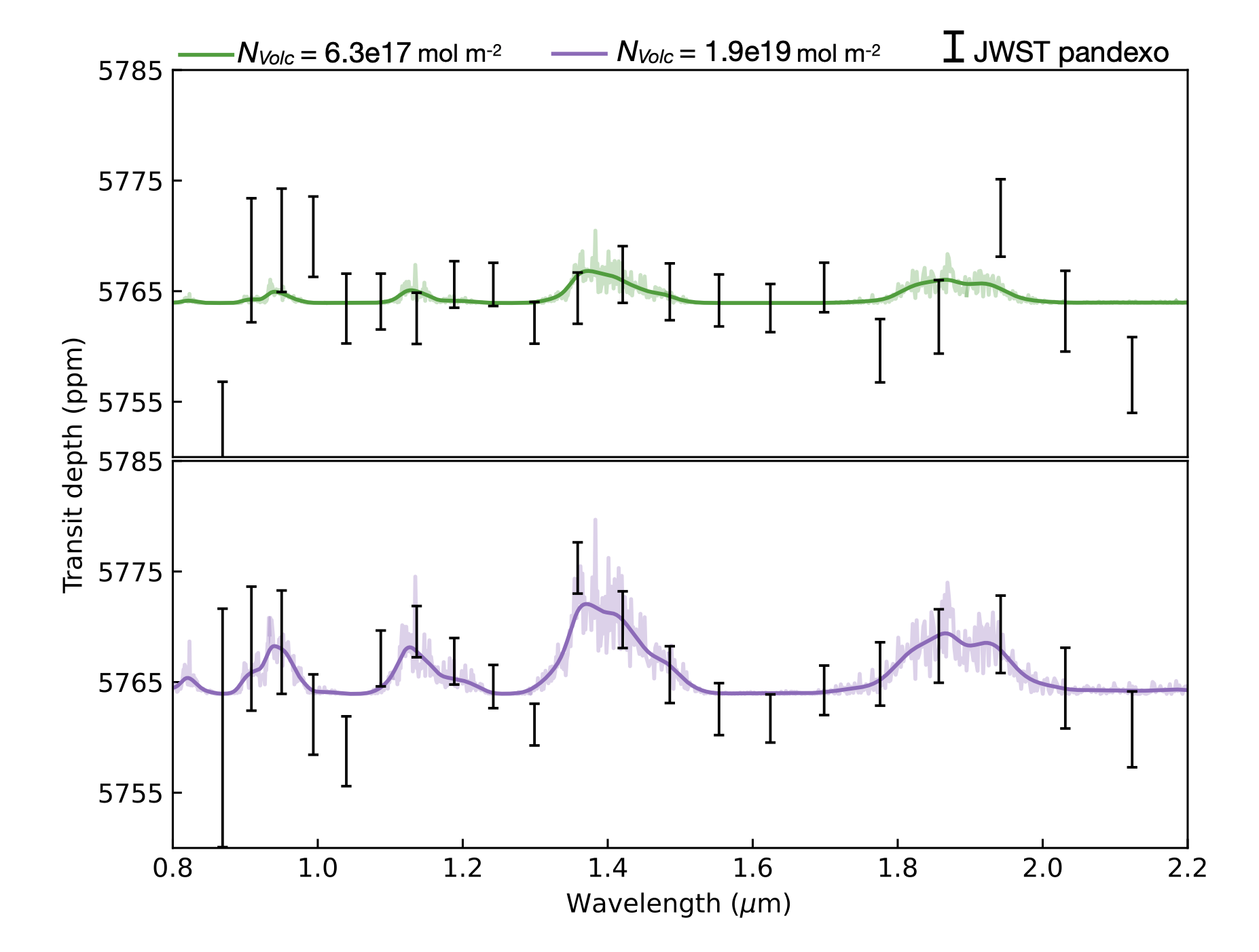}
        \caption{}
    \end{subfigure}
      \vspace{0.2cm}
      
    \begin{subfigure}{0.99\columnwidth}
        \centering
        \includegraphics[width=\linewidth]{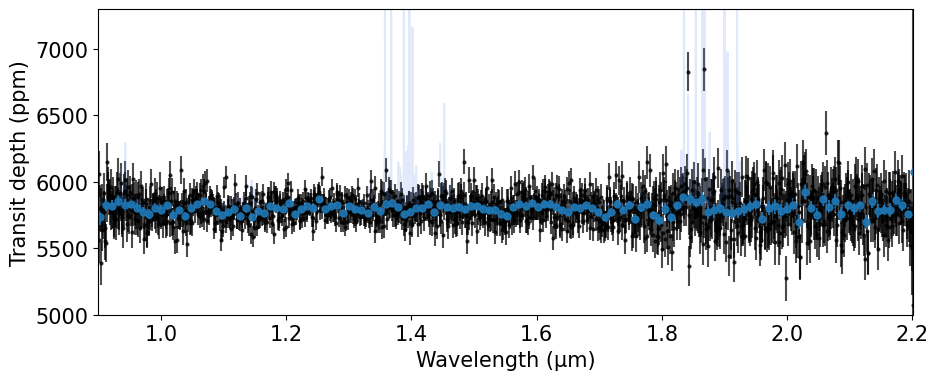}
        \caption{}
    \end{subfigure}

    \vspace{0.2cm}

    \begin{subfigure}{0.99\columnwidth}
        \centering
        \includegraphics[width=\linewidth]{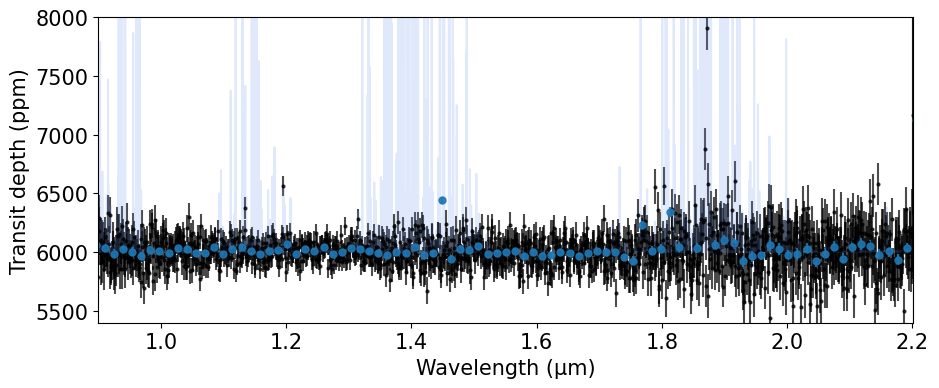}
        \caption{}
    \end{subfigure}

   \caption{ Synthetic transmission spectra of cryovolcanic water outgassing on TRAPPIST-1f, taking into account cryovolcanism, sublimation and volatile loss from XUV irradiation. The black data points are generated with \textit{PandExo} for JWST/NIRISS.  (a)  Spectrum of a tenuous hydrostatic water atmosphere for $N_{total} = 1.9 \times 10^{19}$ (hydrostatic benchmark) produced using \textit{petitRADTRANS} (purple line). Mock JWST datapoints assume 450 transits and a binned resolution of 12. (b) Spectrum of a uniformly distributed sputtered cloud produced using \textsc{Prometheus} (blue line). The darker blue data point are binned observations, demonstrating how the signal at $\approx$ 1.9 microns gets diluted. This sub-figure corresponds to a $N_{total} = 1.1 \times 10^{24}$ molecules $m^{-2}$ (low sublimation case). Mock JWST datapoints assume 20 transits. (c) Same as (b) but for $N_{total} = 1.6 \times 10^{26}$ molecules $m^{-2}$ which corresponds to the benchmark of the high sublimation case. Mock JWST datapoints assume 15 transits.} 
    \label{fig:spectrum}
\end{figure}

Figure~\ref{fig:spectrum} presents model transmission spectra of cryovolcanic outgassing on TRAPPIST-1~f, generated using the radiative transfer codes and exposure time calculator described in Section~\ref{sec:cryovolcanism_methods}. The spectra are shown for two representative column densities and are compared to synthetic JWST/NIRISS data (black error bars). The top panel illustrates a tenuous hydrostatic atmosphere and serves as a detectability benchmark, while the middle and bottom panels correspond to a non-hydrostatic, globally uniform sputtered cloud.
If JWST were to observe every transit of TRAPPIST-1~f ($P = 9.2$ days) over a 12-year mission lifetime, corresponding to approximately 450 transits, a hydrostatic water column density of $N_{\mathrm{volc}} \simeq 1.9\times10^{19} molecules  \,\mathrm{m^{-2}}$ would become marginally detectable, as the 1.4\,$\mu$m water feature exceeds the synthetic JWST uncertainties. 
In contrast, for a globally uniform sputtered cloud, detectability requires a substantially larger total volatile inventory of $n_{\mathrm{tot}} \sim 10^{38}$ molecules, corresponding to an effective column density of $N_{\mathrm{volc}} \sim 10^{23}molecules\, \mathrm{m^{-2}}$.

\begin{figure}[!htb]
    \centering
    \includegraphics[width=0.99\columnwidth]{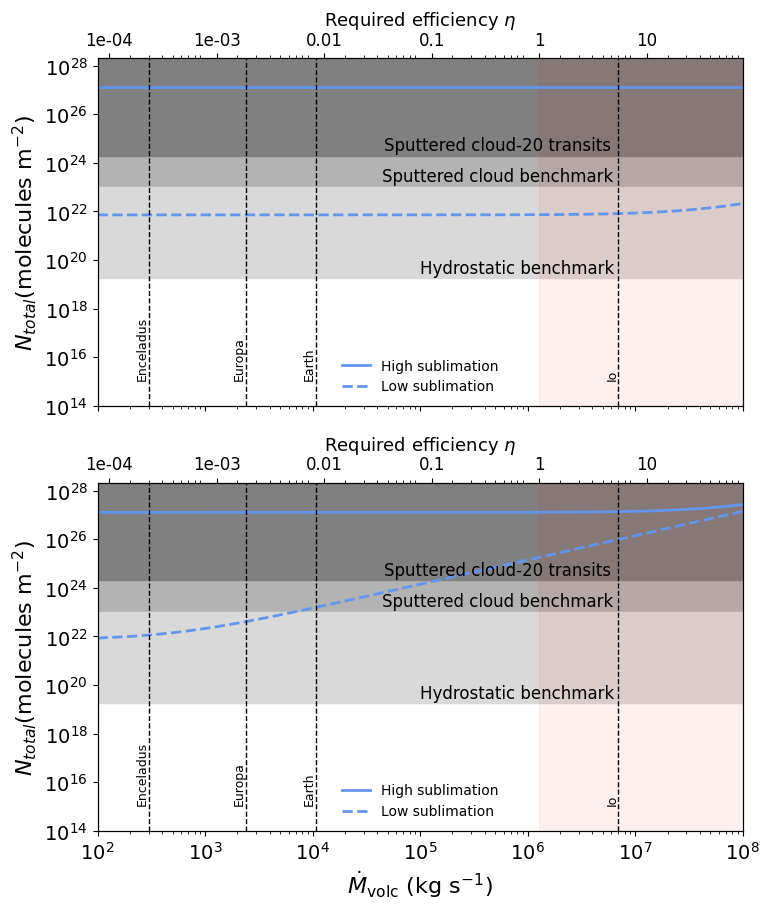}
   \caption{ $N_{total}$ as a function of $\dot{M}_{volc}$ for TRAPPIST-1f. The shaded red region indicates parameter space requiring efficiencies $\eta>1$, which are unphysical. The hydrostatic and sputtered cloud benchmarks represent the column densities observable with JWST/NIRISS with 450 transits for a hydrostatic atmosphere and a collisionless exosphere, respectively. The darkest grey region indicates an exosphere's column density observable with 20 transits. Top: Detectability of water column densities for a globally uniform sputtered cloud.  Even with 450 transits, a collisionless exosphere cannot get observed for low-sublimation. Bottom:  Detectability of water column densities assuming the sputtered material is confined to a surface fraction $f = 0.004$, yielding enhanced line-of-sight columns, which represent plume-like configurations. Any detection at column densities that would otherwise require $\eta > 1$ therefore, would imply that the absorbing material is spatially confined rather than globally distributed, consistent with localized sputtering or plume-like activity or high water sublimation.}
    \label{fig:nvolc}
\end{figure}

To evaluate whether the column densities inferred from the outgassing models presented in Section \ref{sec:cryovolcanism_methods} are physically and observationally plausible, Figure~\ref{fig:nvolc} shows the total water column density, $N_{\mathrm{total}}$, as a function of the volcanic mass loss rate, $\dot{M}_{\mathrm{volc}}$, and the corresponding efficiency factor $\eta$ for TRAPPIST-1~f. Vertical dashed lines indicate representative volcanic mass loss rates inferred for Enceladus, Europa, Earth, and Io. The light grey shaded region indicates the column densities required for detectability based on hydrostatic atmosphere spectra (Figure \ref{fig:spectrum}a), while the darker grey region corresponds to the detectability benchmarks derived from non-hydrostatic, sputtered-cloud spectra (Figure \ref{fig:spectrum}b - Sputtered cloud benchmark).

The top panel corresponds to a globally uniform  $N_{\mathrm{total}}$, which represents a conservative detectability scenario for non-hydrostatic exospheres ($f=1$). Even when assuming an observing campaign of 450 transits with JWST, the column densities required for detection fall within a regime that demands efficiencies $\eta>1$, indicating that a spatially uniform sputtered cloud is energetically unphysical under the assumptions of our low-sublimation model (dashed blue line).The shaded red region highlights this unphysical parameter space. In contrast, the high-sublimation case (continuous blue line) lies well above the JWST detectability threshold corresponding to $\sim 20$ transits.This suggests that the signal could be detectable even under the conservative assumption of a globally uniform exosphere (f=1) and for physically realistic efficiency factors. A detection in this regime could thus provide quantitative constraints on the sublimation rate.

The bottom panel shows the same models under the assumption that the sputtered material is spatially confined to a surface fraction $f = 0.004$, yielding plume-like configurations that enhance the effective line-of-sight column density (Equation \ref{eq:Ntot_loc}).  Such localization allows the low-sublimation modeled columns to approach values higher than the hydrostatic detectability benchmark. Because these configurations are still modeled as non-hydrostatic, collisionless exospheres, their detectability is more consistently assessed relative to the sputtered-cloud benchmark. When evaluated against this more appropriate benchmark, localized sputtering represents a substantially more favorable scenario than a globally uniform cloud. For $\dot{M}_{\mathrm{volc}}$ between Earth-like and Io-like values, a region of observability for low-sublimation emerges that requires 20 transits. We note that this outcome depends additionally on the assumed efficiency $\eta$, which quantifies how effectively internal energy is converted into volcanic outgassing and remains poorly constrained. High sublimation column densities are observable for both uniformly distributed and localized sputtered clouds with $< 20$ transits.

Taken together, these results show that (i) hydrostatic water atmospheres provide an upper bound on the detectability of cryovolcanic water vapor, (ii) globally uniform sputtered exospheres with low sublimation rates are energetically inconsistent with detectable transmission signals, and (iii) spatial localization of outgassed material reduces the energetic requirements of a detection (iv) a detection would provide quantitative constraints on the sublimation rate of an icy exoplanet. These findings place constraints on the detectability of cryovolcanic water vapor on TRAPPIST-1~f with JWST and motivate future observations with next-generation facilities.

Cryovolcanic and sputtered neutral environments in the Solar System provide useful context for interpreting the column densities and geometries considered here. In a simple way, Europa hosts either a sputter-dominated water exosphere or a sublimation-dominated exosphere in which neutrals follow largely ballistic trajectories, producing extended clouds of neutral water dissociation products with column densities of order $10^{12}$-$10^{14}\mathrm{molecules}$ $\mathrm{cm^{-2}}$, depending on the orbital phase  \citep{OZA201923}. For instance, it is well known the darker trailing hemisphere of Europa is far hotter and also coincident with plasma ram \citep{Lorenz}. Ganymede, by contrast, possesses an intrinsic magnetosphere that enhances ion-neutral interactions resulting in more compact distributions following Ganymede's field lines \citep{LEBLANC2023115557}. Shallow profiles ($q \sim 3$) imply extended, Europa-like distributions, whereas steeper profiles ($q \sim 6$), as used in this work, confine the same inventory closer to the planet, more analogous to plasma-modified environments such Ganymede's magnetic exosphere. 

 For comparison with Solar System literature, the water geysers at Enceladus were probed by JWST to have a total of $\mathcal{N}_{tot} \sim $ 10$^{34.9}$ water molecules \citep{geronimo23}, roughly a third which begin orbiting Saturn \citep{CJ2010}. In the scenario of a high fraction of atmospheric escape then, a circumstellar water torus may also be amenable for detection about the ultra cool dwarf star. A detection of a tenuous circumstellar torus would be indicative of a cryovolcanic system venting directly to space \citep{geronimo23}. A larger, total inventory of $n_{\mathrm{tot}} \sim 10^{38}$ water molecules corresponds to an average stellar-disk line-of-sight (LOS) column density of $\langle N \rangle \approx 1\times10^{15.6} \mathrm{molecules}$ 
$\mathrm{cm^{-2}}$ when normalized by the stellar disk area, following the definition of $\langle N \rangle$ used in \citet{Oza_2019}. This value is comparable than columns discussed for neutral column densities at Europa \citep{OZA201923, Roth2014} and Ganymede's water exosphere \citep{LEBLANC2023115557}.

\section{Discussion}\label{sec:discussion}
In this section, we place our modeling results in the context of previous studies, discuss the detectability of cryovolcanic activity on TRAPPIST-1f, g, and h, and outline directions for future observational and theoretical work.

\subsection{Water Mass Fraction comparison with previous work}
\begin{figure}[!htb]
    \centering
    \includegraphics[width=0.99\columnwidth]{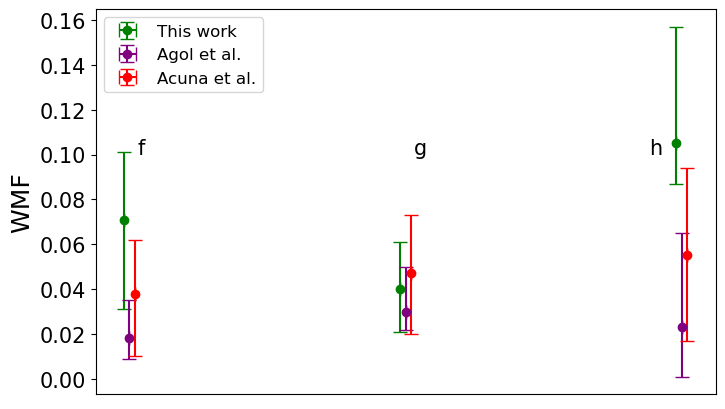}
    \caption{Estimated water mass fraction, WMF together with their uncertanties for Trappist-1 f,g,h with our model (green), \citet[][ purple]{Agol_2021}, and \citet[][ red]{acuna2021}.}
    \label{fig:WMF_comparison}
\end{figure}

In addition to refining the mass/radius measurements for the Trappist-1 planets, \citet{Agol_2021} theorized over their possible interior structures and water content. They estimated their WMFs assuming a condensed water layer of 300 K at 1 bar.  \citet{acuna2021} also modelled the interior and hydrospheres of the Trappist-1 planets, taking into account Equations of State (EOS) for different phases of water, including water vapor and high pressure ice phases. In Figure~\ref{fig:WMF_comparison}, we show the calculated WMFs for CMF = 0.25 from our model, taking into account the mass and radius uncertainties (Table \ref{table:planet_parameters}), and we compare them to those derived by \citet{Agol_2021,acuna2021}. We compare our results solely with the latter two studies, as previous work \citep{Barr_2018, Dorn_2018} on the interior structures of the system have not used the updated values for mass and radius. In particular, they have modelled the interior of the Trappist-1 planets using the mass and radius measurements, constrained by \citet{delrez_2018, grimm_2018}. Our results (Figure \ref{fig:WMF_comparison}) are mostly within the uncertainties of \citet{acuna2021}, and there is a larger discrepancy with the results of \citet{Agol_2021}, whose results are significantly lower (for planets f and h). This discrepancy can be explained by our different modelling strategies.  In order to reach the given density constraints, dictated by the mass and radius measurements, \citet{Agol_2021} used a condensed water layer, without taking into account any equation of state for different phases, thus assuming lower density for the water layer. The agreement of our results with \citet{acuna2021} is to be expected, as we included similar phases for water (high pressure ice polymorphs). Our results can be interpreted as one out of the multiple interior structures that \citet{acuna2021} predict, since their error-bars are calculated for CMF values that span between 0.2 and 0.3, and we only show the interior structure for CMF = 0.25.

\subsection{Water content in the Trappist-1 system and other possible interior structures} \label{subsection: water_content_formation_scenario}


 The extended Pre-Main-Sequence (PMS) phase of M-dwarf stars, like Trappist-1, can cause atmospheric mass loss via thermal escape, and  water loss  due to the increased luminosity and XUV activity \citep[see e.g. ][]{Wordsworth_2018, Luger_2015}. This water mass loss depends on the planet's orbital distance \citep{BourrierTrappist17}, making close-in planets more prone to volatile escape \citep{bolmont2017}.
\citet{gialluca2024implications} calculated the water loss rates for the Trappist-1 planets, and studied their implications on recent observations \citep{Zieba_2023,Greene_2023}, which imply a tenuous oxygen atmosphere or no atmosphere for planets b and c, respectively. According to \citet{gialluca2024implications}, however, the loss of atmospheres on the inner planets does not preclude the retention of substantial water inventories on the outer planets, where lower stellar irradiation could allow water to persist. Recent studies have assessed the complete loss of a water reservoir at the Galilean satellites, namely Io  \citep{Bennacer2026} and Europa, both which experience atmospheric sputtering \citep{Johnson1990}, or plasma-driven escape of sodium and potassium \citep{Gebek2020}. The sputter escape of water has yet to be modeled in detail at Europa and Ganymede \citep{OZA201923, LEBLANC2023115557}, in contrast to Io's observed sodium escape by atmospheric sputtering \citep{Burger2001}, however is to first order, photoionization limited.

As seen from Figure \ref{fig:nvolc}, tidal heating rates and, thus $N_{volc}$ is highly dependant on the CMF that is assumed for the planets. We explore the range of probable CMFs for the planets that are compatible with their formation mechanism. 
The most favored scenario for the Trappist-1 planets is formation in a protoplanetary disk, possibly beyond the water ice line, with a subsequent inward migration, which is supported by the resonant chain and compactness of the system \citep{Teyssandier2022}.
Planet formation within a protoplanetary disk results in planets with similar ratios of relative refractory elements, meaning similar CMFs, for fully differentiated bodies \citep{Bond_2010, ELSER2012859}, but with different volatile element content \citep{Oberg_2016}. Thus, the WMF of the planets is also a function of their formation mechanism.
If the Trappist-1 planets are best described by the CMF of the most dense planet (planet c with $22\% - 31\%$ CMF), their densities can be fitted with different light-element content or water, according to  \citet{Dorn_2018}.
With our model, the latter CMF ratios would mean WMF values of $\approx4-8 \times 10^{-2}, 3-6 \times 10^{-2}$, and $1-11 \times 10^{-2}$ for Trappist-1f,g and h respectively.

%
%
Varying the WMF in order to obtain the observed densities is
compatible with the results of \citet[][ Figure \ref{fig:WMF_comparison}]{Agol_2021}, who found that all Trappist-1 planets fall within one compositional line that is less dense than solar system terrestrial planets.
Other possible compositions for the Trappist-1 planets include either an interior with less iron compared to the terrestial content, or oxidized iron in the mantle \citep{Agol_2021}.
%

\subsection{Future work}

A first question to address in the future would be how to confirm that an observation of water comes from cryovolcanism. Exogenic sources of water are expected to remain constant over time \citep{papagini2020}, whereas endogenic cryovolcanism may vary depending on the balance of interior processes such as tidal heating and radiogenic heat generation. Repeated observations could potentially contribute to the confirmation of the origin of water. The relationship between cryovolcanism and the different interior heat sources remains an open question, with significant implications for interpreting observations of cryovolcanic activity. 

A second point of discussion would be a possible connection of observations of cryovolcanic activity to interior constraints. If tidal heating is the dominant driver of cryovolcanism, planets with low CMF values are more likely to exhibit detectable cryovolcanic signatures due to enhanced tidal dissipation. Conversely, if the total internal heat budget is dominated by radiogenic sources, planets with higher CMF values might exhibit stronger cryovolcanic activity.

 Further investigation is needed to determine feasible interior models for a particular water column density observation. A more detailed analysis would need to consider uncertainties in interior parameters and ice rheology, the temperature dependence of rheological properties and the identification of secondary equilibrium states in mantle dynamics, as in \citet{Kleisioti2023}. While our study outlines the potential sources of cryovolcanism and their implications, the detailed uncertainties inherent to interior modeling and observational constraints fall outside the scope of this work. Future research is needed to refine these models and better constrain the interplay between planetary structure, heat generation, and cryovolcanic activity.

\section{Conclusions}\label{sec:conclusions}

In this work, we combined thermal and tidal heating models, previously applied to icy satellites in the Solar System on TRAPPIST-1f, g, and h. We identified interior configurations compatible with the presence of subsurface oceans and showed that, for all three planets, tidal heating within the high-pressure ice layer and radiogenic heating dominate over tidal dissipation in the silicate mantle and ice I shell under the assumptions of our model. Our dissipation partitioning differs from Solar System icy satellites, where dissipation is often concentrated in the ice~I shell, because our TRAPPIST-1 solutions can contain very thick viscoelastic HPI layers which represent a large volume of deformable solid and therefore contribute substantially to the total tidal dissipation budget.

In addition, we constrain the ice I and subsurface ocean layer thicknesses for all three planets.
The interior structures of the inner two planets, Trappist-1f and g, would be in thermal equilibrium only in cases where the outer ice I shell is relatively thin, between $\approx$ 4 and 10 km, under the assumptions of our model.
We also calculated the corresponding WMFs and CMFs that are compatible with such interior structures. We find that for an Earth-like CMF (0.32) Trappist-1f,g,h, would possess a subsurface ocean in thermal equilibrium for a WMF of $8.5 \times10^{-2}, 5.9 \times10^{-2}$ and $11.8\times10^{-1}$. 

Moreover, we investigated the detectability of cryovolcanic water vapor via transmission spectroscopy. We showed that hydrostatic water atmospheres provide an upper bound on detectability, whereas non-hydrostatic, globally uniform sputtered exospheres require energetically unphysical efficiencies to produce detectable transmission signatures with JWST when pure water ice sublimation rates are low. Spatial localization of outgassed material in plume-like configurations substantially enhances effective line-of-sight column densities and reduces the energetic requirements for a detection. Such configurations are detectable on TRAPPIST-1f for volcanic mass fluxes between Earth-like and Io-like values with $\approx$ 20 transits, although such detections depend sensitively on the assumed energy conversion efficiencies and spatial localization of outgassed material. When high water ice sublimation rates are assumed for TRAPPIST-1f non-hydrostatic, globally uniform and localized sputtered clouds produce signals above the JWST/NIRISS detectability threshold corresponding to $\approx$20 transits, under the assumptions of our models. Since the high sublimation rate is standard for pure water ice \citep{bob81}, our simulations suggest a potential detection (or non-detection) of water vapor is able to inform the pure water mass fraction of ice on TRAPPIST-1f's surface.

Taken together, our results place constraints on the observability of cryovolcanic water vapor on TRAPPIST-1f, g, and h with current facilities. While JWST observations need favorable conditions to detect cryovolcanic activity, future observations with next-generation space- (e.g. \textit{Habitable World Observatory}) and ground-based telescopes (e.g. ELT/ANDES), combined with improved modeling of interior–surface interactions, will be essential for assessing the prevalence and observational signatures of cryovolcanism on icy exoplanets \citep{Quick_2023}.

\begin{acknowledgements}

The authors would like to thank the anonymous reviewers and St\'ephanie Cazaux for their constructive reviews which greatly strengthened the manuscript.
The {\tt Python} programming language \citep{rossum1995} and the open-source {\tt Python} packages {\tt numpy} \citep{walt2011}, {\tt matplotlib} \citep{hunter2007}, {\tt astropy} \citep{astropy2013}, \textit{petitRADTRANS} \citep{prt2019,prt2020,prt2022} and \textit{PandExo} \citep{pandexo} were used. For the calculation of tidal heating the LOV3D software
package \citep{Rovira_Navarro_2024} was used. This research has been supported by the PEPSci Programme (Planetary and ExoPlanetary Science Programme), NWO, the Netherlands.

\end{acknowledgements}
\bibliographystyle{aa}

\bibliography{source}

\newpage

\begin{appendix}
\section{Melting temperature as a function of pressure} 

\begin{figure}[!htb]
    \centering
    \includegraphics[width=\columnwidth]{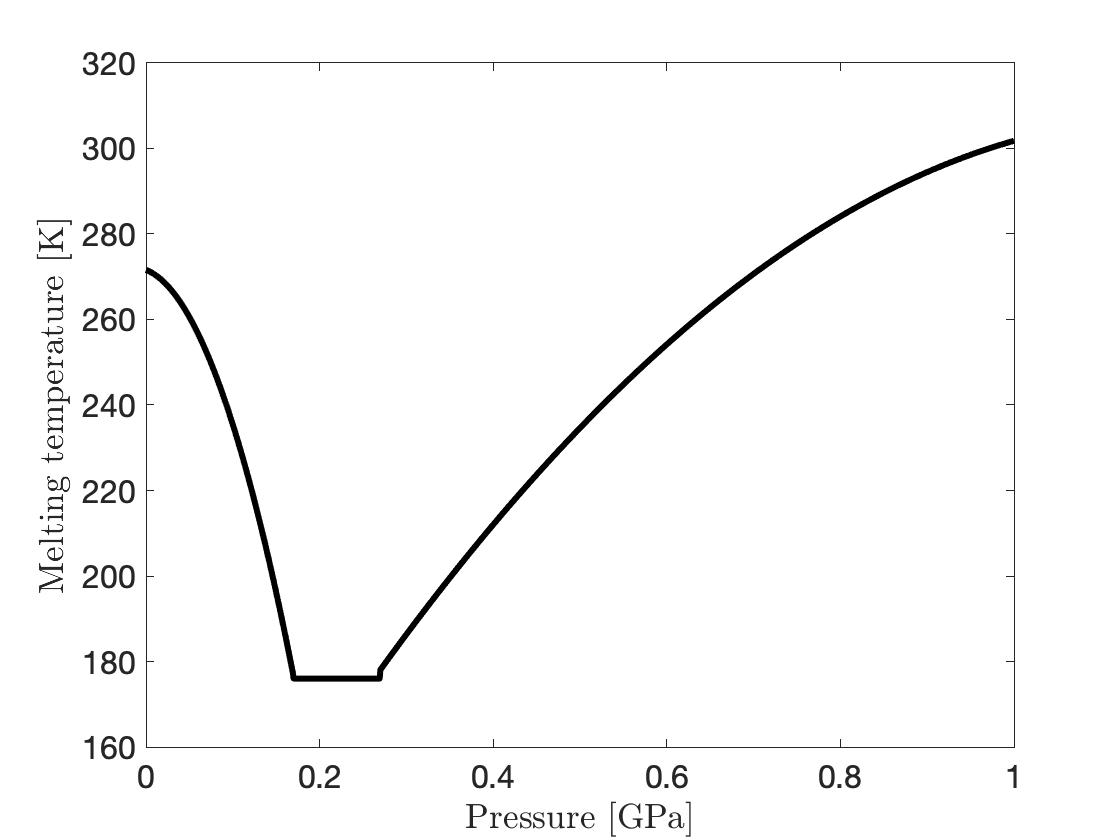}
    \caption{Melting temperature as a function of pressure for an ammonia containing ocean of 5 wt.$\%$ \citep{GRASSET2000617, sohl_2003}}
    \label{fig:meltingTvsP}
\end{figure}
\section{Tidal heat generated in the Ice I outer shell} 

\begin{figure}[!htb]
    \centering
    \includegraphics[width=\columnwidth]{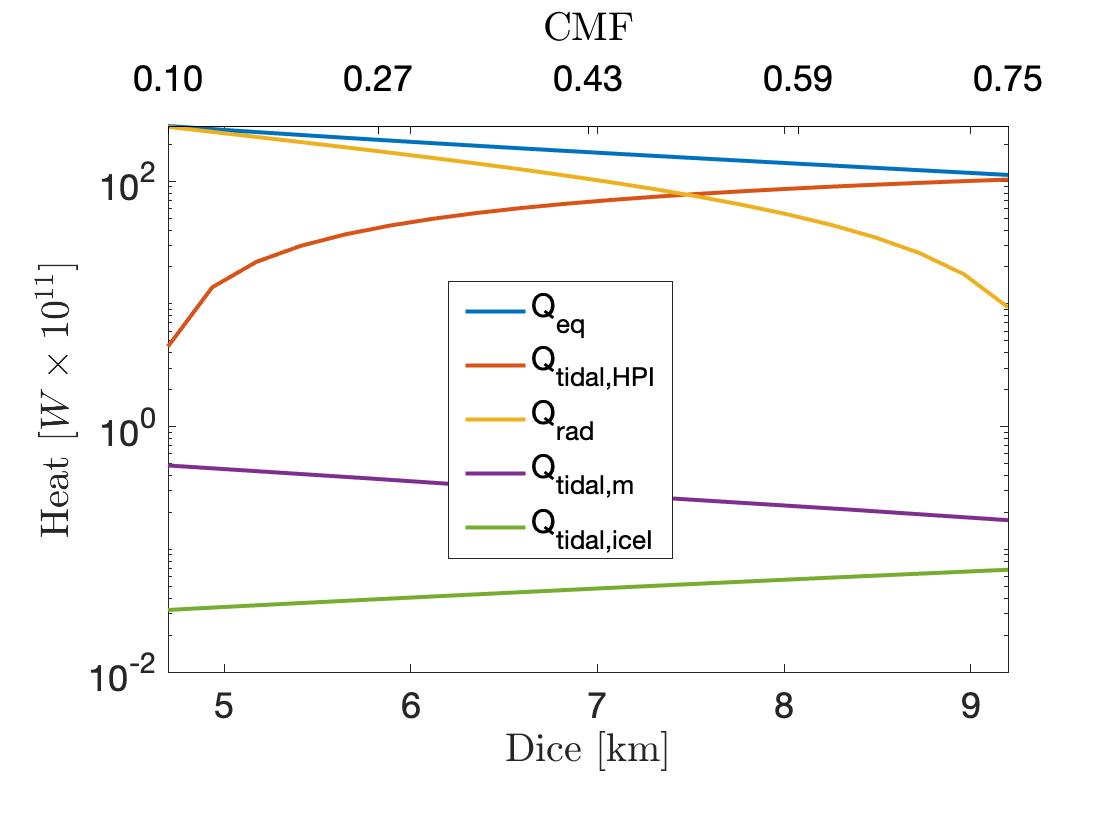}
    \caption{Heat source distribution in the interior of Trappist-1f. $Q_{tidal,HPI}$ $Q_{tidal,m}$, and $Q_{tidal,iceI}$ is the tidal heat generated in the HPI layer, mantle and ice I shell respectively. $Q_{rad}$ is the radiogenic heat produced in the mantle. The tidal heat that is generated in the ice I layer at least 4 orders of magnitude smaller than the total heat generated in the interior ($Q_{eq}$).}
    \label{fig:Tidalheatice1}
\end{figure}

\section{Interior structures as a function of \texorpdfstring{$D_{ice}$}{D ice} for the planets Trappist-1g,h} 

\begin{figure}[!htb]
    \centering
    \includegraphics[width=\columnwidth]{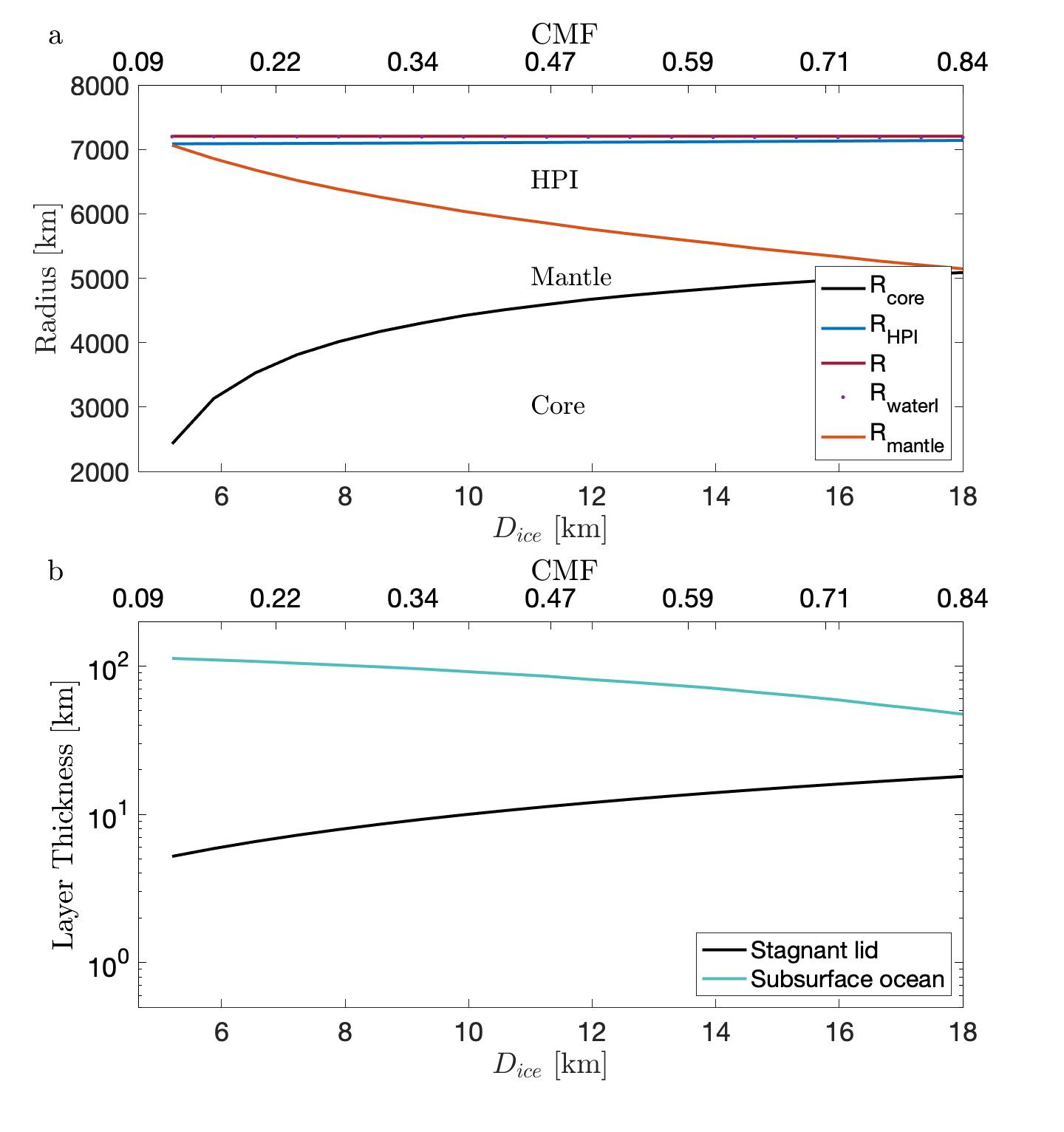}
    \caption{a. Interior structure as a function of $D_{ice}$ for the planet Trappist-1g.}
    \label{fig:interiorg}
\end{figure}

\begin{figure}[!htb]
    \centering
    \includegraphics[width=\columnwidth]{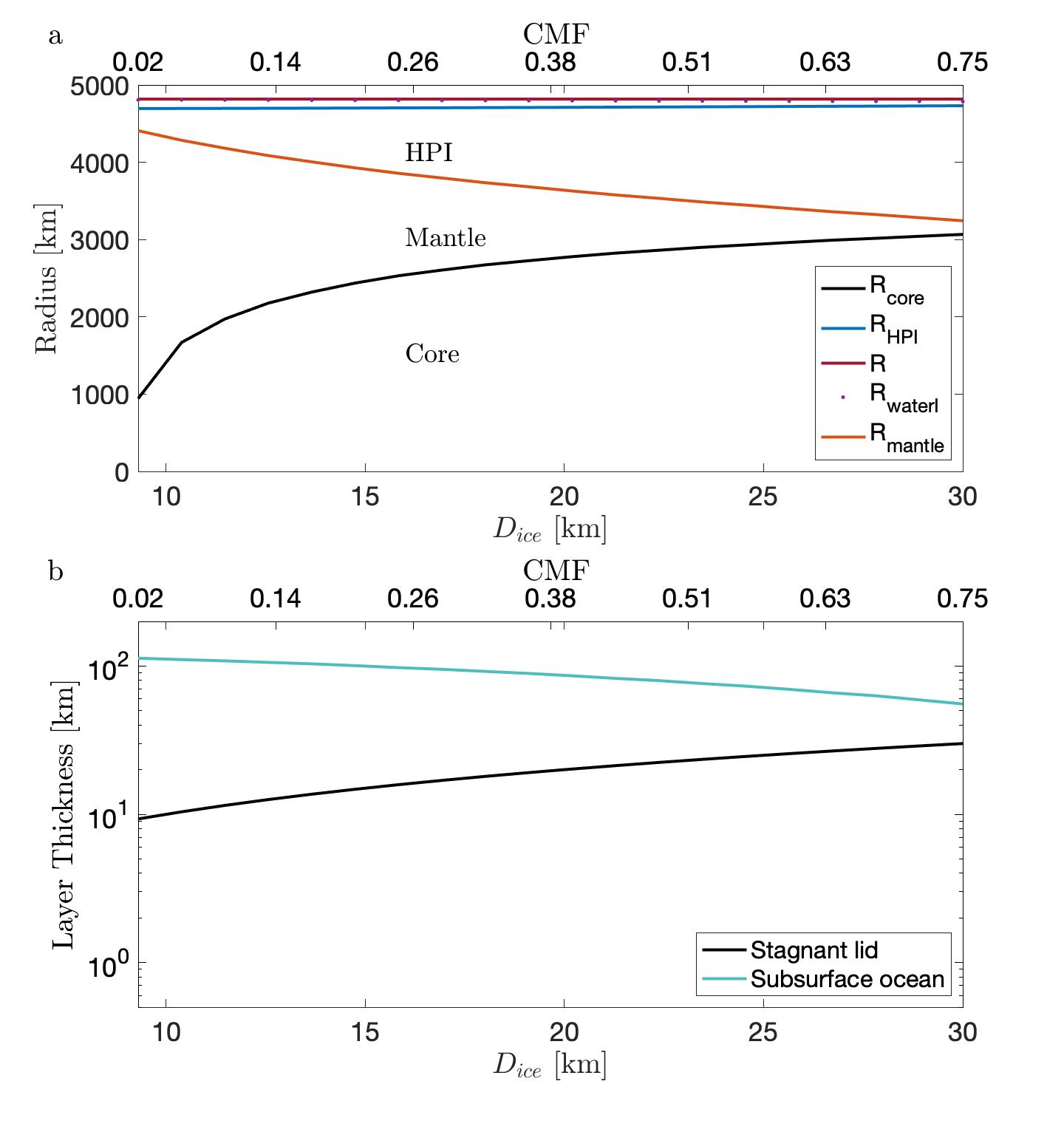}
    \caption{(a) Interior structure as a function of $D_{ice}$ for the planet Trappist-1h.}
    \label{fig:interiorh}
\end{figure}
\end{appendix}

\end{document}